\documentclass[structabstract]{aa}

\usepackage{graphicx}
\usepackage{txfonts, longtable, dcolumn, ctable}
\usepackage{natbib}
\bibpunct{(}{)}{;}{a}{}{,}

\begin{document}

   \title{A study of asteroid pole-latitude distribution based on an extended
set of shape models derived by the lightcurve inversion method}


   \author{J. Hanu{\v s}
          \inst{1*}
          \and
          J. {\v D}urech\inst{1}
          \and
          M. Bro{\v z}\inst{1}
          \and
          B. D. Warner\inst{2}
          \and
          F.~Pilcher\inst{3}
           \and
	  R. Stephens\inst{4}
           \and
	  J. Oey\inst{5}
	   \and
          L. Bernasconi\inst{6}
           \and
          S. Casulli\inst{7}
           \and
	  R. Behrend\inst{8}
	   \and
	  D. Polishook\inst{9}
           \and
	  T. Henych\inst{10}
           \and
	  M. Lehk\'y\inst{11}
           \and
	  F. Yoshida\inst{12}
           \and
	  T. Ito\inst{12}
   }

   \institute{Astronomical Institute, Faculty of Mathematics and Physics,
Charles University in Prague,
              V~Hole{\v s}ovi{\v c}k{\'a}ch 2, 18000 Prague, Czech Republic\\
              $^*$\email{hanus.home@gmail.com}
         \and
             Palmer Divide Observatory, 17995 Bakers Farm Rd., Colorado Springs,
CO 80908, USA
	 \and
	     4438 Organ Mesa Loop, Las Cruces, NM 88011, USA
	 \and
	     Goat Mountain Astronomical Research Station, 11355 Mount Johnson
Court, Rancho Cucamonga, CA 91737, USA
	 \and    
	     Kingsgrove, NSW, Australia
	 \and
	     Observatoire des Engarouines, 84570 Mallemort-du-Comtat, France
	 \and    
	     Via M.~Rosa, 1, 00012 Colleverde di Guidonia, Rome, Italy
	 \and    
	     Geneva Observatory, CH-1290 Sauverny, Switzerland
	 \and
	     Benoziyo Center for Astrophysics, The Weizmann Institute of
Science, Rehovot 76100, Israel
	 \and
	     Astronomical Institute, Academy of Sciences of the Czech Republic,
Frièova 1, CZ-25165 Ondøejov, Czech Republic
	 \and
	     Severni 765, CZ-50003 Hradec Kralove, Czech republic
	 \and
	     National Astronomical Observatory, Osawa 2-21-1, Mitaka, Tokyo
181-8588, Japan
  }

   \date{Received 17-02-2011 / Accepted 13-04-2011}

 
  \abstract
   {In the past decade, more than one hundred asteroid models were derived using
the lightcurve inversion method. Measured by the number of derived models,
lightcurve inversion has become the leading method for asteroid shape
determination.}
   {Tens of thousands of sparse-in-time lightcurves from astrometric projects
are publicly available. We investigate these data and use them in the lightcurve
inversion method to derive new asteroid models. By having a greater number of models with
known physical properties, we can gain a better insight into the nature of
individual objects and into the whole asteroid population.}
   {We use sparse photometry from selected observatories from the AstDyS
database (Asteroids -- Dynamic Site), either alone or in combination with dense
lightcurves, to determine new asteroid models by the lightcurve inversion method.
We investigate various correlations between several asteroid parameters and
characteristics such as the rotational state and diameter or family membership. We
focus on the distribution of ecliptic latitudes of pole directions. We create a
synthetic uniform distribution of latitudes, compute the method bias, and compare the 
results with the distribution of known models. We also construct a
model for the long-term evolution of spins.}
   {We present 80 new asteroid models derived from combined data sets where
sparse photometry is taken from the AstDyS database and dense lightcurves are
from the Uppsala Asteroid Photometric Catalogue (UAPC) and from several
individual observers. For 18 asteroids, we present updated shape solutions based
on new photometric data. For another 30 asteroids we present their partial models, i.e., 
an accurate period value and an estimate of the ecliptic latitude of the pole. The addition
of new models increases the total number of models derived by the
lightcurve inversion method to $\sim$200. We also present a simple statistical
analysis of physical properties of asteroids where we look for possible
correlations between various physical parameters with an emphasis on the spin
vector. We present the observed and de-biased distributions of ecliptic latitudes
with respect to different size ranges of asteroids as well as a simple 
theoretical model of the latitude distribution and then compare its predictions with
the observed distributions. From this analysis we find that the latitude distribution of small asteroids ($D <
30$ km) is clustered towards ecliptic poles and can be explained by the YORP
thermal effect while the latitude distribution of larger asteroids ($D > 60$ km)
exhibits an evident excess of prograde rotators, probably of primordial origin.}
   {}
 

  \titlerunning{A study of asteroid pole-latitude distribution}
   \maketitle

\section{Introduction}\label{introduction}

The lightcurve inversion method (LI) is a powerful tool that allows us to derive
basic physical properties of asteroids (the rotational state and the shape) from
their disk-integrated photometry
\citep[see][]{Kaasalainen2001a,Kaasalainen2001b, Kaasalainen2002a}. This
photometry can be \textit{dense-in-time}, which typically consists of tens to a few
hundreds of individual data points observed during one revolution. This is in contrast
to \textit{sparse-in-time}, where the typical separation of individual measurements
is large compared to the rotation period. For sparse data, we usually have a few
measurements per night, such as in the case of astrometric sky surveys. In the
following text, we use the terms ``dense lightcurves'' and ``sparse
lightcurves''.

To obtain a unique spin and shape solution, we need a set of at least a few tens
of dense lightcurves observed during at least three apparitions.
Based on simulated photometric data and the survey cadence of the Panoramic Survey
Telescope and Rapid Response System (Pan-STARRS), \citet{Kaasalainen2004} showed that
we can also use only sparse data for the
inversion technique. In this case, a unique model can be derived from more than about
one hundred
calibrated measurements observed during 3--5 years if the photometric accuracy is
better than $\sim$5\% \citep{Durech2005,Durech2007}. Sparse data available
so far are not that accurate. Nevertheless, for many asteroids with high
lightcurve amplitudes, it is possible to derive their models from current sparse
data (see Section \ref{photometry} for more details). We can also combine sparse
and dense data to derive models. First results from this approach were presented by
\citet{Durech2009}, where sparse data from the US Naval Observatory in Flagstaff (USNO) were used.

Currently (January 2011), there are 113 models of asteroids derived by the
lightcurve inversion method; most of them are stored in the Database of
Asteroid Models from Inversion Techniques
\citep[DAMIT\footnote{\texttt{http://astro.troja.mff.cuni.cz/projects/asteroids3D}},][]{Durech2010}. Most of these
models were derived from dense lightcurves. Only 24 of them were computed
from combined dense and sparse data \citep{Durech2009}. The AstDyS database
(Asteroids -- Dynamic Site\footnote{\texttt{http://hamilton.dm.unipi.it/}}),
which contains data from astrometric projects, is another possible source of
sparse data. However, most of the data are not accurate enough to be used for
inversion alone. On the other hand, even noisy sparse data in combination with a
few dense lightcurves can give us, in many cases, a unique solution
\citep{Durech2007}. The aim of our work was to gather these data, keep only those that
were useful, and then combine them with dense lightcurves in the lightcurve
inversion method.

Dense data are best used to define the rotational period and constrain the
period interval that must be searched during the model computation (see Section
\ref{models} for more details). On the other hand, sparse data usually cover a
long time interval, typically over several apparitions, and carry information about
brightness variations for different geometries, which constrains the pole
directions.

A priori information about rotational periods of asteroids plays an important
role in the process of model determination. When an approximate period is known,
we search for the solution near this value (details in Section \ref{models}) and
thus save considerable computational time. We use the latest update of the Minor
Planet Lightcurve
Database\footnote{\texttt{http://cfa-www.harvard.edu/iau/lists/Lightcurve\-Dat.html}} published by
\citet{Warner2009} to check for previously derived periods. For many asteroids,
there are only a few sparse lightcurves
from different astrometric observatories available but no dense lightcurves. In these cases, we
must scan the whole interval of expected period values (2--30~hours). This
approach is time-consuming and there is no guarantee that the correct period will
lie in the scanned interval.

The knowledge of rotational states of asteroids is fundamental for understanding
the history of the Solar System, specifically the accretion of planets or the collisional
processes. For example, it was presumed that due to collisional evolution,
the spin-vector distribution of Main Belt Asteroids (MBAs) should be nearly
isotropic, possibly with a small excess of prograde spins \citep{Davis1989}.
\citet{Johansen2010} performed a hydrodynamical simulation of the accretion of
pebbles and rocks onto protoplanets and speculated that the trend of prograde
rotators among the largest asteroids is primordial.

First statistical analyses of the spin-vector distribution were presented by
\citet{Magnusson1986,Magnusson1990} and \citet{Drummond1988b,Drummond1991},
later by \citet{Pravec2002} and \citet{Skoglov2002}. They all observed a lack of
poles close to the ecliptic plane. \citet{Kryszczynska2007} used more objects in
the analysis, finding that the distribution was strongly anisotropic with a moderate excess
of prograde spins in the limited size range from 100 to 150 km. Interpretation
of this depopulation of poles close to the ecliptic plane is still unclear.
Probable candidates are selection effects, the role of inclination, the YORP
effect\footnote{Yarkovsky--O'Keefe--Radzievskii--Paddack effect, a torque caused
by the recoil force from anisotropic thermal emission}
\citep{Rubincam2000,Vokrouhlicky2003}, or a combination of these. The YORP effect acts
only on small bodies with $D \lesssim 40$~km. Asteroids with these sizes have
non-Maxwellian spin rate distribution \citep{Pravec2000} and is particularly evident for
asteroids with $D < 14$~km \citep{Warner2009}. It is believed that the YORP effect is responsible
for this trend since it can either spin up or spin down an irregularly-shaped asteroid on
the timescale shorter than the typical time between collisions and also affects
the obliquity of spin axes \citep{Rubincam2000, Bottke2006}.

In the Near Earth Asteroids (NEAs) population, the latitude distribution of
poles is different from that of MBAs \citep{LaSpina2004, Kryszczynska2007}, i.e.,
there is a significant excess of retrograde spins probably caused by the
transport mechanism of MBAs to Earth-crossing space by gravitational resonances
and the Yarkovsky effect\footnote{a thermal force acting on a rotating asteroid}
\citep{Morbidelli2003}. There is no statistically significant clustering in the
longitude of poles of either MBAs or NEAs \citep{Kryszczynska2007}.

As the number of asteroid models with known physical properties grows, we can have a better insight into
the nature of individual objects and into the asteroid population as a whole.
In Section \ref{photometry}, we describe available dense and sparse photometric
data and the selection of astrometric observatories with quality sparse data. In
Section \ref{results}, we present new asteroid models derived from combined
photometric data sets or from sparse data alone, mentioning a few individual
objects and define several procedures on how to test the reliability of new models.
In Section \ref{statistics}, we present a statistical analysis of asteroid
physical parameters that we derived using the lightcurve inversion method or
adopted from different sources (proper elements from the AstDyS database,
diameters from IRAS, \dots). We also present results of a numerical simulation
that allowed us to estimate the bias in pole directions of the lightcurve
inversion method. Using these results, we then corrected the observed pole distributions for this effect.
Finally, in order to explain the observed latitude distributions, we present a
simple theoretical model of the latitude distribution in Section \ref{YORP_sim}.

\section{Photometric data}\label{photometry}

The main source of dense photometric lightcurves is the Uppsala Asteroid
Photometric Catalogue \citep[UAPC, ][]{Lagerkvist1987, Piironen2001}, where the
lightcurves for about 1\,000 asteroids are stored. We also used data from
several individual observers (Table \ref{references}).

Sparse photometry was first used in combination with dense data for lightcurve inversion by \citet{Durech2009}. These sparse data were from the USNO-Flagstaff station and had a typical photometric uncertainty of $\sim$8--10\%. Other sparse photometric
measurements are produced by many astrometric surveys, but mostly as a by-product. In most cases, asteroid magnitudes are
given to only one decimal place, i.e., the accuracy is 0.1 mag at best. Whether or nor this is sufficient for a unique shape determination for reasonable number
of asteroids can be deduced from asteroids lightcurve amplitude distribution.
We used lightcurve amplitude data for $\sim$2\,500 asteroids from the Minor
Planet Lightcurve Database \citep{Warner2009} and found that the mean lightcurve
amplitude is $\sim$0.3 mag. For 19\% of asteroids, the amplitude is $\ge$0.5
mag. This means that, in principle, photometry with an accuracy of $\sim$0.1 mag
carries sufficient information about rotational states and shapes for a significant number
of asteroids.

Our goal was to find out which observatories produce photometry
suitable for lightcurve inversion and to use these data for determining new asteroid models. Through to September 2009 (the time of the data download), data for more
than 350\,000 objects from almost 1\,500 observatories were archived on the
AstDyS server. Some of the observatories contributed with only a few data points,
while others contributed tens of thousands of photometric measurements (e.g., large sky surveys such as the Catalina Sky Survey, LONEOS, or Siding Spring Survey).

\subsection{Data reduction}\label{reduction}

The quality of the sparse photometry archived on the AstDyS varies
significantly. We investigated the photometry carefully by establishing
criteria for its quality. Then, using those criteria, we choose only those data
that were useful for inversion.

For each observatory, we extracted photometric data for the first 10\,000
numbered asteroids if there were at least 30 data points for a single
lightcurve. We then transformed this photometry to the standard format used in
lightcurve inversion: we computed geometry of observation (astrocentric ecliptic
coordinates of the Sun and the Earth), corrected for light-time, normalized the
brightness to the distance of 1~AU from the Sun and the Earth, and excluded
clear outliers. 

\begin{figure}
	\begin{center}
	 \resizebox{\hsize}{!}{\includegraphics{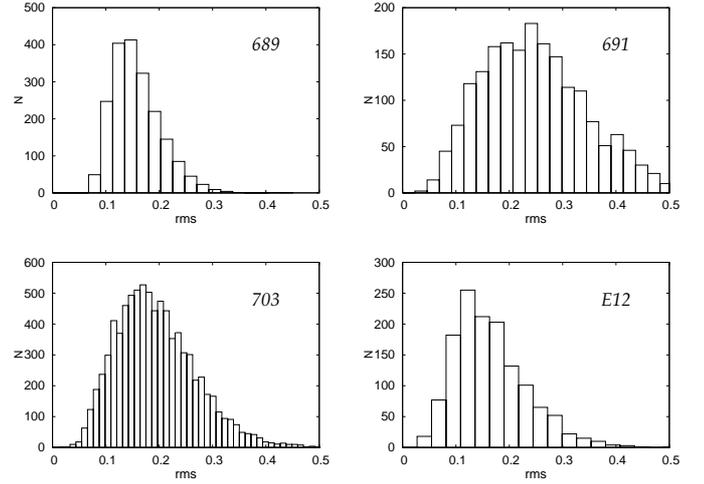}}\\
	\end{center}
	\caption{\label{rms_distribution}Four histograms of residuals comparing actual data against the model given by Eq.\ref{phase_curve} of all sparse lightcurves
belonging to the following observatories: 689 USNO, 691 Steward Observatory, 703
Catalina Sky Survey and E12 Siding Spring Survey. Number of bins is
$\sim\sqrt{N}$, where $N$ is the total number of sparse lightcurves used for
histogram construction.}
\end{figure}

For further investigation, we selected 13 observatories that fulfilled the
condition of having data for more than $\sim$50 asteroids. This resulted in almost
30\,000 sparse lightcurves for $\sim$9\,000 asteroids. In the next step, we
estimated mean uncertainties of individual observatories and, based on these
uncertainties, we assigned a relative weight to the data from each observatory. In this process, we assumed that the brightness vs. solar phase angle relation can be fitted with a simple relation for each sparse lightcurve:

\begin{equation}\label{phase_curve}
 f(\alpha)=\mathrm{cos}^2\left(\frac{\alpha}{2}\right) \left[a\:\exp\left(
{-\frac{\alpha}{b}}\right) +c\:\alpha+d\right],
\end{equation}
where $\alpha$ is the solar phase angle\footnote{the Sun--asteroid--Earth angle}
and $a, b, c$ and $d$ are free parameters. Then, we constructed a histogram of
residuals (rms) for each observatory comparing actual data against the model given
by Eq.~\ref{phase_curve}. Four examples are plotted in
Fig.~\ref{rms_distribution}. The dispersion is caused by observational
uncertainties and by the amplitudes of the lightcurves. From these histograms,
we estimated the ``FWHM''\footnote{the width of the distribution in
the half of its maximum} values and the most frequent residual (the mode); median values of the residual distributions for each location (Table~\ref{observatories}) were computed. Observatories with a high median or ``FWHM''
value ($\gtrsim$0.2 mag for both) are not suitable
for the lightcurve inversion (e.g. observatory 691 in
Fig.~\ref{rms_distribution}). Data from only seven observatories, listed in
Table~\ref{observatories} with non-zero weights, had sufficient accuracy and so could be used for modeling. Based on the values of medians and
``FWHMs'', we estimated a weight for the photometric data from each observatory relative to dense data, which has a unity weight (see
Table~\ref{observatories}). We assumed that the typical accuracy of dense
lightcurves is $\sim$0.02 mag. 

\begin{table*}\caption{\label{observatories}Comparison of estimated
characteristics of residuals for 13 selected observatories: mode, ``FWHM'' and median.}
\begin{center}
 \begin{tabular}{rrccc r ll}\hline
  Obs & $N_{\mathrm{LC}}$ & \multicolumn{1}{c} {Mode} & \multicolumn{1}{c}
{FWHM} & \multicolumn{1}{c} {Median} & \multicolumn{1}{c}
{$\overline{N_{\mathrm{P}}}$} & \multicolumn{1}{c} {Weight} & Observatory name\\
\hline\hline
  608 & 2459 & 0.26 & 0.20 & 0.27 & 37 & 0 & Haleakala-AMOS\\
  644 & 2567 & 0.22 & 0.22 & 0.24 & 36 & 0 & Palomar Mountain/NEAT\\ 
  683 & 218 & 0.18 & 0.25 & 0.20 & 39 & 0 & Goodricke-Pigott Observatory,
Tucson\\
  689 & 1970 & 0.14 & 0.12 & 0.15 & 118 & 0.3 & U.S. Naval Observatory,
Flagstaff\\
  691 & 1893 & 0.23 & 0.22 & 0.24 & 39 & 0 & Steward Observatory, Kitt
Peak-Spacewatch\\
  699 & 546 & 0.17 & 0.11 & 0.18 & 33 & 0.1 & Lowell Observatory-LONEOS\\
  703 & 8350 & 0.17 & 0.16 & 0.19 & 54 & 0.15 & Catalina Sky Survey\\
  704 & 8333 & 0.42 & 0.17 & 0.42 & 311 & 0 & Lincoln Laboratory ETS, New
Mexico\\
  950 & 80 & 0.14 & 0.11 & 0.15 & 180 & 0.15 & La Palma\\
  E12 & 1354 & 0.14 & 0.16 & 0.15 & 41 & 0.1 & Siding Spring Survey\\
  G96 & 1810 & 0.14 & 0.22 & 0.17 & 43 & 0.1 & Mt. Lemmon Survey\\
  H07 & 161 & 0.20 & 0.18 & 0.23 & 47 & 0 & 7300 Observatory, Cloudcroft\\ 
  Hip & 49 & 0.10 & 0.10 & 0.11 & 53 & 0.3 & Hipparcos satellite\\ \hline
 \end{tabular} 
\tablefoot{
For each observatory, the table gives also the number of sparse lightcurves $N_{\mathrm{LC}}$, and the average number of data points for a single lightcurve $\overline{N_{\mathrm{P}}}$.
}
\end{center}
\end{table*} 

The USNO in Flagstaff (MPC code 689) and the Hipparcos satellite are clearly the
best observatories with respect to photometric accuracy. Other observatories are
less accurate but they still hold some information about rotational states and
shapes of asteroids. For any given asteroid, we have typically 2--4 sparse lightcurves
from different observatories covering the last $\sim$10--15 years.
Data from USNO were already used by \citet{Durech2009}. We updated those data along with
adding another 20--30\% of new data from the years 2008 and 2009 if there was an apparition for the asteroid.

\section{Results}\label{results}

\subsection{Models of asteroids}\label{models}

Our lightcurve inversion (LI) method is based on the optimization of unknown
parameters of the shape (modeled as a {\em convex} hull), the rotational state, and the
scattering law \citep[see][]{Kaasalainen2001a,Kaasalainen2001b}. The parameter
space have many local minims. Since LI is a gradient-based method that converges
to a local minimum near the initial choice of parameter values, it is critical to find the global minimum for the data set and then do the modeling. Finding a global minimum involves a systematic search through all relevant parameter values.
Each model corresponding to a particular local minimum is characterized by a
single value $\chi^2$, which corresponds to the quality of the fit. 

A unique solution is defined as follows: (i) the best period has at least 10\%
lower $\chi^2$ than all other periods in the scanned interval, (ii) for this
period, there is only one pole solution with at least 10\% lower $\chi^2$ than
the others (with a possible ambiguity for $\lambda \pm 180^{\circ}$), and
(iii) this solution fulfills our additional tests (see Section
\ref{tests}).

The most time-consuming part of the lightcurve inversion method is scanning
through all periods within a chosen interval, which we determined by
using the periods and reliability codes given in the Minor Planet Lightcurve
Database\footnote{\texttt{
http://cfa-www.harvard.edu/iau/lists/Lightcurve\-Dat.html}, there is also an
explanation and more details about the reliability codes too}. For each asteroid, we scanned an
interval centered at the reported period value $P$ with a range of
$\pm 1\%$, $\pm 5\%$ and $\pm 20\%$ of $P$ for reliability codes 4, 3 and 2,
respectively. Half and double periods were tested later on.

We combined relative lightcurves from the UAPC and from individual observers
together with sparse data obtained from the AstDyS site to create a data set for each asteroid. This gave us data sets for
$\sim$2\,300 asteroids (in $\sim$900 cases there were only sparse data
available) to which we then applied the lightcurve inversion method and then we ran
the additional tests described in Section~\ref{tests}. We derived 80 new unique
models, 16 of which are based only on sparse data. Basic characteristics of
these models are listed in Table~\ref{models_tab}. We estimated the uncertainty
in the pole direction as $\pm$10--20$^{\circ}$ based on previous results with
limited data sets. As might be expected, the uncertainty seems dependent on the number of dense and
sparse photometric data. The longitude uncertainty increases for
higher latitudes because meridians on a ($\lambda$, $\beta$)-sphere are more
dense with increasing latitude. These uncertainties are discussed in more detail in
Section~\ref{simulations}. The uncertainty of the rotational period depends on
the time interval covered by observational data and is of the order of the last
decimal place of period values $P$ in Tables~\ref{models_tab},
\ref{partials_tab} and \ref{preliminary}.

In some cases, we were able to determine a unique rotational period, but we had
multiple pole solutions with similar ecliptic latitudes $\beta$. These models
give us accurate period values and rough estimates of ecliptic latitudes
$\beta$, which are also important parameters. In Table~\ref{partials_tab}, we
present results for 30 partial models, where $\beta$ is the mean value for all
different models if the dispersion is smaller than $50^{\circ}$. We defined a
parameter $\Delta=|\beta_{\mathrm{max}}-\beta_{\mathrm{min}}|/2$ as being the
estimated uncertainty of $\beta$, where $\beta_{\mathrm{max}}$ and
$\beta_{\mathrm{min}}$ are the extremal values within all $\beta$. 

All new unique shape models are now included in DAMIT.

\subsection{Comments to selected models}\label{selected_objects}

In DAMIT there are several solutions designated as ``preliminary''. These
models do not have a well-constrained pole solution or are based on combined data
sets. For 18 of those asteroids we derived updated model solutions based on additional
photometric data (see Table \ref{preliminary}). The difference between the old and new model for asteroid (1223) Neckar was significant.
The new model has a slightly different period, but the pole
directions and shapes are nearly similar to the old model. The current data suggest a
period of $P=7.82401$ hours (previous value was $P=7.82123$ hours).

The asteroid (4483) Petofi was recently observed by Brian Warner. We derived a
shape solution from three poor dense lightcurves and one sparse lightcurve from
Catalina Sky Survey. Warner used these four lightcurves in combination
with his new observations and also derived the shape model of Petofi
\citep{Warner2011}. His period of $P=4.3330$ hours and pole direction
(90$^{\circ}$, $35^{\circ}$) are close to our solution of $P=4.33299$ hours
and (107$^{\circ}$, $40^{\circ}$).

The asteroid (832) Karin was also studied by \citet{Slivan2010}; their solution
with $P=18.352$ hours and pole (51$^{\circ}$ or 228$^{\circ}$, $41^{\circ}$)
confirms our results, see Table \ref{models_tab}.

In past decades, occultations of stars by several asteroids were observed. These
events give us additional information about the shape and can help resolve
which mirror solutions of a model is the correct one. According to the recent work of
\citet{Durech2011s}, asteroid occultation measurements prefer pole solutions
of (122$^{\circ}$, $-44^{\circ}$) for (10) Hygiea, (347$^{\circ}$, 47$^{\circ}$)
for (152) Atala, (28$^{\circ}$, $-72^{\circ}$) for (302) Clarissa,
(223$^{\circ}$, 67$^{\circ}$) for (471) Papagena, and (296$^{\circ}$,
41$^{\circ}$) for (925) Alphonsina. Spin solutions preferred by asteroid
occultations appear in bold font in Tables \ref{models_tab} and \ref{preliminary}.

\subsection{Models and method testing }\label{tests}

We constructed five additional tests to be sure that the new models are
reliable. We performed the first two tests for all models. For models derived
only from sparse data, which are presented for the first time, we performed three additional tests:

\paragraph{Inertia tensor of the shape model.}

The lightcurve inversion method we use assumes that asteroids are in a relaxed
rotational state, which means that derived models should rotate around the axis
with a maximum moment of inertia. For each derived shape, we computed principal
moments using equations presented by \citet{Dobrovolskis1996} and checked if
the rotation axis was close to the principal axis of the maximum momentum of
inertia. We rejected those models for which the angle between the spin axis and
the axis with a maximum momentum of inertia was larger than $\pm$30$^{\circ}$.
However, this criterion is too strict for elongated models with similar sizes
along the rotational axis and the axis that is both perpendicular to the rotational
axis and is the minimal size of the model. In this case, the principal moments for
these two axes are similar. Under these circumstances, the angle between the spin
axis and the axis with a maximum momentum of inertia can be large even for
realistic shapes and so we allowed these models to pass this test.

\paragraph{Half- and double-period models.}

In cases where we have only a few dense lightcurves for a given asteroid, it is
easy to confuse the correct rotational period with its half or double value. When the a
priori period value in the Minor Planet
Lightcurve Database is uncertain, which corresponds to
a low reliability code (see Section~\ref{models}), it is
reasonable to check if the half and double period value give a better
fit. If the period was in doubt, we searched for a solution also around $2P$ and $P/2$;  if the $\chi^2$ was lower than $1.1\chi^2$ of the solution with period $P$, we
rejected the model as unreliable.

\paragraph{Reduction of the number of sparse photometric data.}
In this test, we used only the sparse data sets for modeling. For 63 asteroids, this led to unique shape
and spin state solutions (after performing the tests described above). For each solution, we randomly reduced the original amount of
observed sparse data points to 90\% and used these new limited data sets again
in the lightcurve inversion. Our expectation was that we would not get a
unique solution when using less data. This was true for five asteroids. These
models, when using the full sparse data sets, are not necessarily wrong, but the amount of available data is probably just at the
level when a unique solution can be derived. The important point of this test is that, for a given asteroid, we did not find two different but formally correct solutions when using the full versus reduced data sets. In Table \ref{models_tab}, we present 16 models successfully derived only from sparse data that passed this test.

\paragraph{Models from sparse data vs. DAMIT.}
Here, we used previously derived models based only on relative photometry which
are stored in DAMIT. As can seen in the previous test, sparse data are sometimes
sufficient to produce a unique model of an asteroid. In 16 cases, we were able
to derive a model for an asteroid which was already included in DAMIT and thus a
model based on entirely different photometric data sets is available. These two
independent models can be then compared and should be similar. We obtained
similar resulting models for all 16 asteroids.

\begin{figure}
	\begin{center}
	 \resizebox{\hsize}{!}{\includegraphics{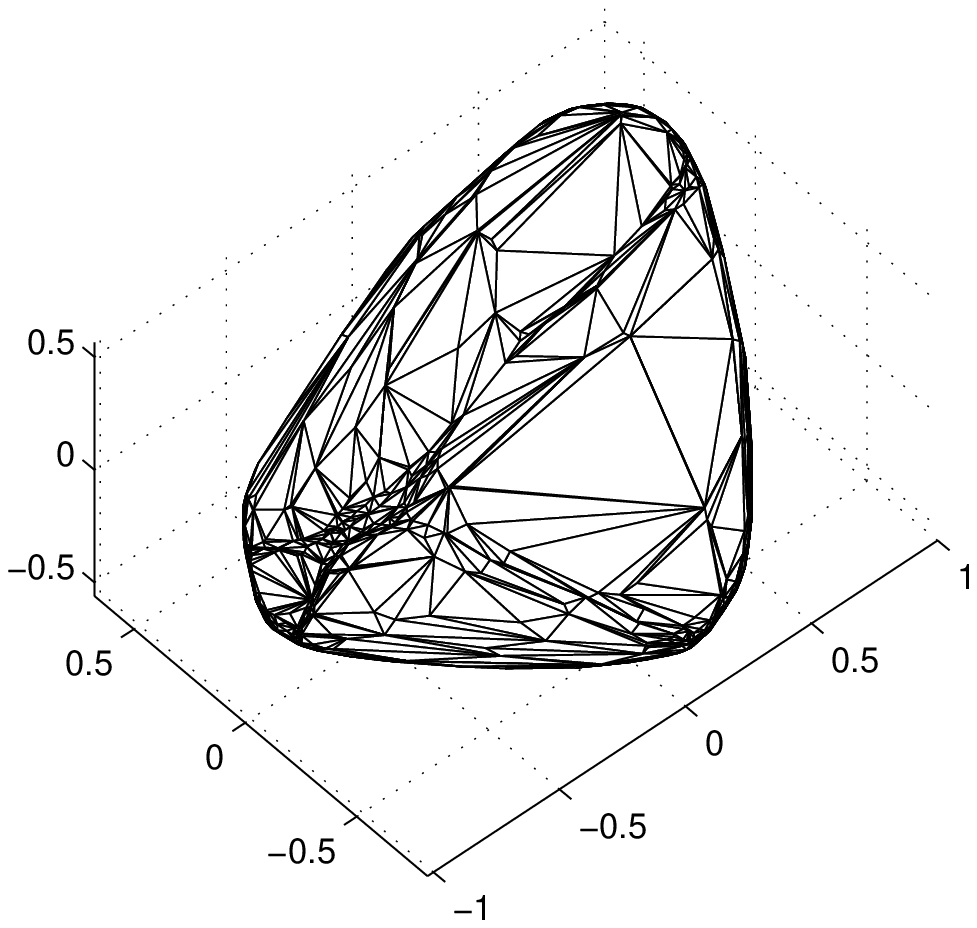}\includegraphics{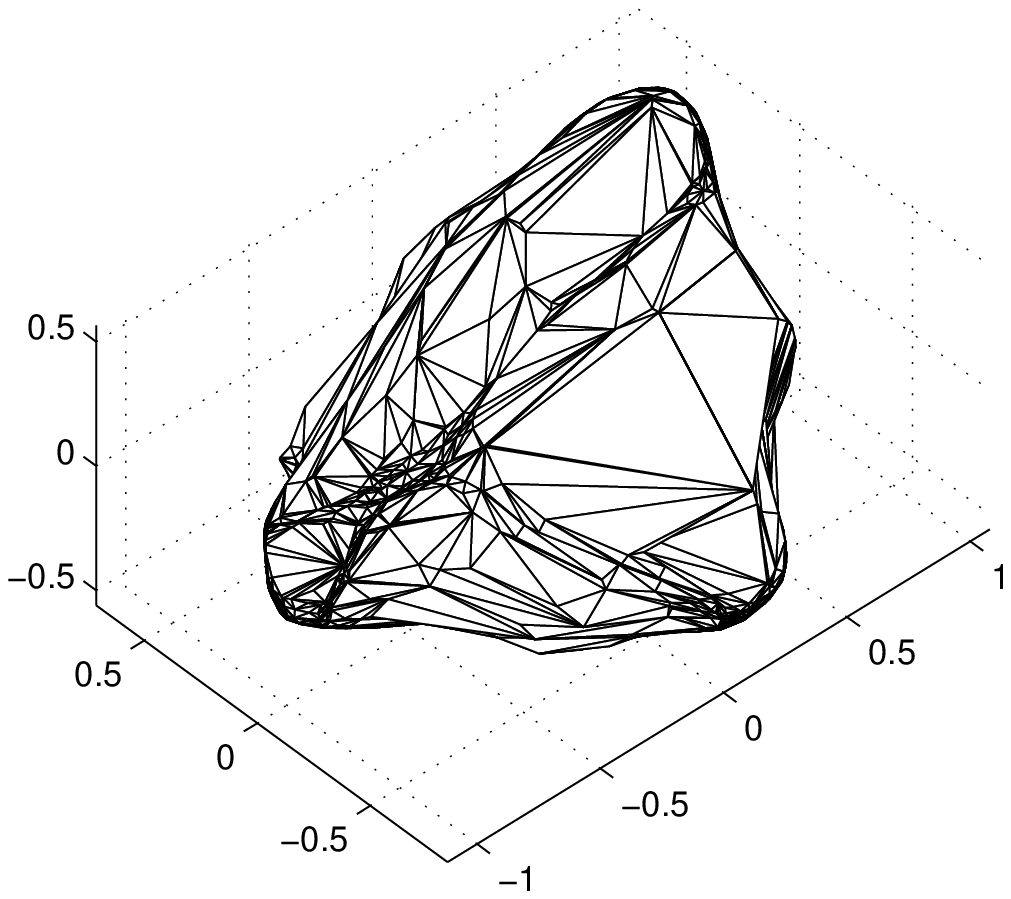}}\\
	\end{center}
	\caption{\label{mock}Asteroid (810) Atossa: shape model (left panel) and an example of a ``mock'' shape model (right panel).}
\end{figure}

\paragraph{Models of ``mock'' objects.}
For each asteroid shape model derived only from sparse data, we created
a set of ten ``mock'' objects of roughly the same appearance and spin state (see
an example of such shape in Fig. \ref{mock}). For these synthetic objects, we
computed their photometric data using same epochs and geometries and with
similar random noise level. These synthetic photometric data sets were then used
in the lightcurve inversion method. A check that the original model using actual data is reliable is to be able to derive most of the
models of the ``mock'' objects. The dispersions between the periods and pole
directions of the ``mock'' objects represent
the typical uncertainties of these parameters. For all studied asteroids, we
were able to derive unique models for most of their ``mock'' objects. In
all cases when we did not get a unique solution for the ``mock'' object, the
best fit corresponded to the correct solution although other solutions could not
be ruled out. The typical uncertainty in pole direction was $\pm$10$^{\circ}$
and, for the period, $\sim$0.1 times the difference between the two local minimums as
determined by the period value and the time span of the data.

\section{An analysis of periods, poles and sizes of asteroid
models}\label{statistics}

Previous studies have been looking on spin state results using different techniques,
e.g. amplitude--epoch, lightcurve inversion or radar methods. If there were multiple
solutions for a given asteroid, the most probable one or
simply the weighted mean was taken. However, this could cause systematic
deviations. In our study, we used only results based on the lightcurve inversion
method -- i.e., unique and partial models presented in this work and models from
the DAMIT. Our sample consists of 221 asteroid models: 80 new models, 18 updates
for models from DAMIT, 30~new partial models, 84~models from DAMIT and 9~new
models presented by \citet{Durech2011s}. Our sample consists of models for 206~MBAs,
10~NEAs, 3~Hungaria, 1~Trojan, and 1~Hilda and so a statistical study is
only possible for the MBAs. In many cases there is an ambiguity in the pole
direction since there are two, undistinguishable mirror solutions. For our
statistical analysis we randomly chose only one.

\begin{figure*}
	\begin{center}
	 \includegraphics[width=17cm]{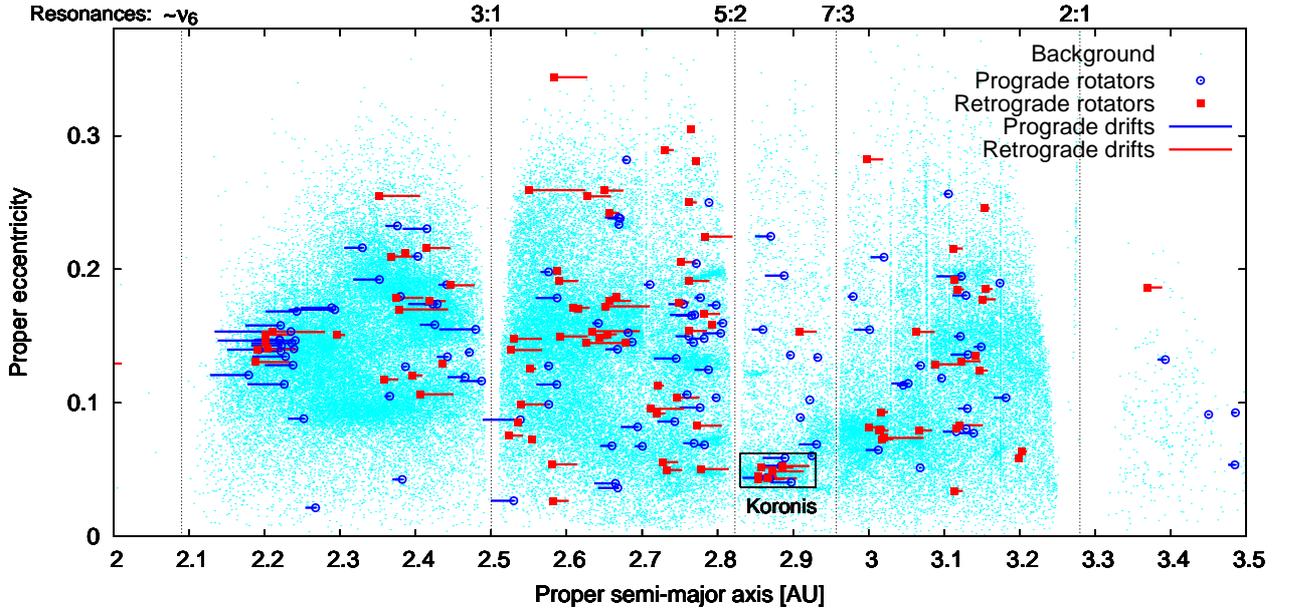}\\
	\end{center}
	\caption{\label{a_vs_e}Relation between the proper semi-major axis and
the proper eccentricity for asteroids in our sample and for the first 100\,000
numbered asteroids for comparison. Main resonances are shown by dotted lines.
Prograde rotators are plotted with blue circles and retrograde rotators with red
squares. The horizontal lines represent for each asteroid its estimated past
drift (i.e. where the asteroid came from) during the collisional lifetime
computed with Eq.~\ref{tau_reor}. Proper elements are from the AstDyS database.}
\end{figure*}

In Fig.~\ref{a_vs_e}, we show (among other things to be discussed later) the
relation between the proper semi-major axis and the proper eccentricity for
asteroids in our sample and for all main belt asteroids. It is obvious that
the positions of studied asteroids strongly correlate with the MBAs population
and so derived models are not significantly biased with respect to orbits, e.g.
they do not lie in the inner main belt. Several asteroids in
Fig.~\ref{a_vs_e} with semi-major axis $a >$ 3.3~AU belong to the
Cybele group, e.g. (121) Hermione.

\subsection{Pole distribution analysis}\label{spin_vector}

\begin{figure}
	\begin{center}
	 \resizebox{\hsize}{!}{\includegraphics{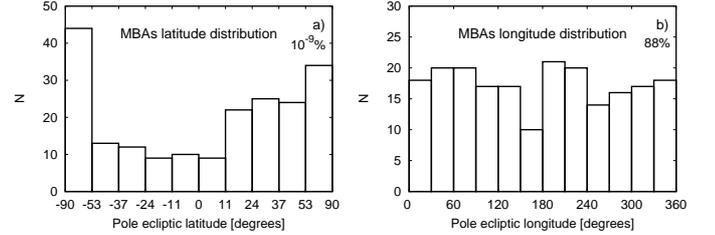}}\\
	\end{center}
	\caption{\label{mbo_pole_distribution}Fig. a) shows the ecliptic
latitude distribution of all MBAs in our sample except the Koronis family
members. The width of bins in latitude $\beta$ corresponds to similar surfaces
on the ($\lambda$, $\beta$)-sphere, so the bins are equidistant in sin$\beta$.
In Fig. b), the longitude distribution of all MBAs is plotted, again except the
Koronis family members. The longitude bins are equidistant in $\lambda$. In the
top right corners, there are the probability values of the $\chi^2$-tests (that
the observed distributions are randomly drown from a uniform distribution, see
Table \ref{chi2}).}
\end{figure}

In the following study of spin axis directions, we did not use the Koronis
family members because their spin states are correlated, i.e., their spin vectors are
clustered towards two values of the obliquity \citep{Slivan2002}. In
Fig.~\ref{mbo_pole_distribution}a, we show the ecliptic latitude distribution of
our MBA sample. As in all similar plots, the width of the latitude and longitude
bins corresponds to equal surfaces on the ($\lambda$,~$\beta$)-sphere (bins are
equidistant in $\sin\beta$ for latitudes and in $\lambda$ for longitudes). We
confirmed the expectation that there is a lack of asteroids with latitudes close
to the ecliptic plane. The latitude distribution is clearly not symmetric: about
half of the retrograde rotators have latitudes in the bin
($-$53$^{\circ}$,~$-$90$^{\circ}$). On the other hand, less than a
third of asteroids with prograde spins are in the corresponding bin
(53$^{\circ}$,~90$^{\circ}$). Moreover, the remaining prograde bins are more
populated than the corresponding retrograde ones. From a detailed look at the
plot we can see that there are up to 10\% more prograde rotators among the MBAs.

\begin{figure}
	\begin{center}
	 \resizebox{\hsize}{!}{\includegraphics{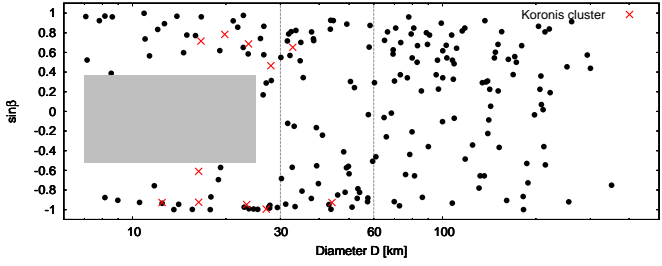}}\\
	\end{center}
	\caption{\label{D_vs_beta}Dependence of the ecliptic latitude $\beta$
(plotted as $\sin\beta$) of the pole direction on the models diameter $D$. The
gray box shows the gap of small latitudes for asteroids with $D < 30$ km.
Members of the Koronis cluster are plotted with crosses.}
\end{figure}

\begin{figure}
	\begin{center}
	
\resizebox{\hsize}{!}{\includegraphics{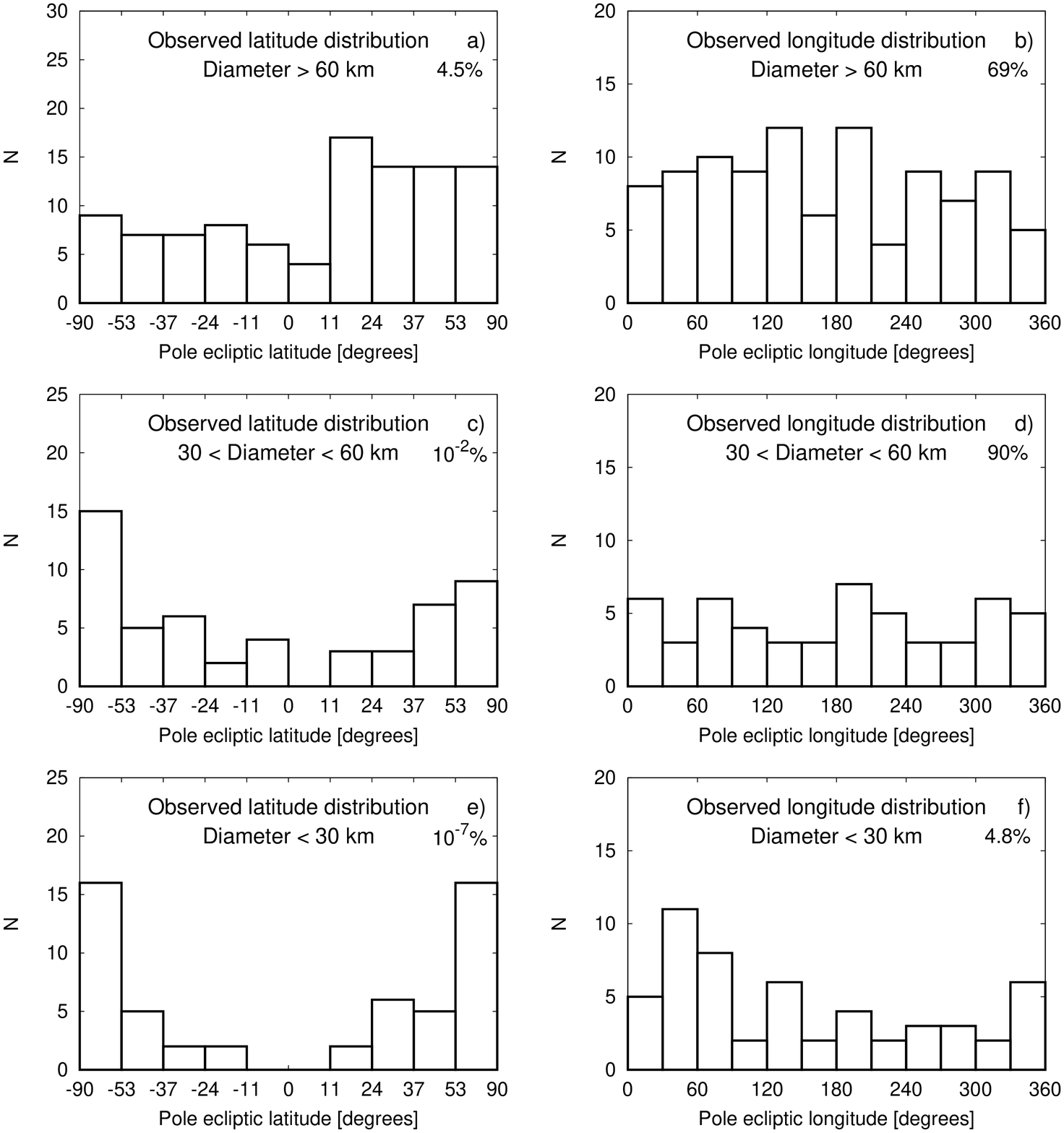}}\\
	\end{center}
	\caption{\label{mbo_pole_distribution_diam}Histograms showing the
observed latitude and longitude distributions of MBAs (except Koronis cluster
members) for different size ranges. Fig. a) shows the latitude distribution for
asteroids with diameters larger than 60 km, c) for asteroids with diameters in
the range of 30--60 km, e) for asteroids with diameters smaller than 30 km, and
similarly b), d) and f) for longitudes. The width of bins in the latitude
$\beta$ and longitude lambda $\lambda$ corresponds to similar surfaces on the
($\lambda$, $\beta$)-sphere, so the bins are equidistant in sin$\beta$ and
$\lambda$. In the top right corners, there are the probability values of the
$\chi^2$--tests (that the observed distributions are randomly drown from a
uniform distribution, see Table \ref{chi2}).}
\end{figure}

In Fig.~\ref{D_vs_beta}, we show the dependence of the ecliptic latitude $\beta$
of the pole direction on the diameter $D$ (most of the diameters used are based
on the IRAS data \citep{Tedesco2002} or occultations profiles and have an uncertainty $\pm$10\%). Even for diameters $D \lesssim 50$~km, the
clustering of the latitudes towards higher absolute values, and conversely, the depletion close to the ecliptic plane is obvious and markedly so for $D\lesssim30$~km. Asteroids with larger diameters have a more isotropic distribution of latitudes and only a moderate excess of prograde. In Fig.~\ref{mbo_pole_distribution_diam}, we plotted the latitude
and longitude distributions of asteroids with respect to their diameters. Based
on Fig.~\ref{D_vs_beta}, we resolved three different size groups: 0--30~km,
30--60~km and $>$60~km.

The latitude distribution for asteroids with $D > 60$~km
(Fig.~\ref{mbo_pole_distribution_diam}a) is close to the uniform distribution
for latitudes lower than 11$^{\circ}$ and for larger latitudes it exhibits an
evident excess of prograde rotators. This is in agreement with theoretical
arguments presented by \citet{Davis1989} and recently by \citet{Johansen2010}.
On the other hand, the latitude distribution for asteroids with $D < 30$~km
(Fig.~\ref{mbo_pole_distribution_diam}e) exhibits a strong depopulation of pole
vectors close to the ecliptic plane (i.e. small absolute values of latitudes
$\beta$). The few asteroids with small latitudes have diameters $D > 25$~km. The
latitude distribution for asteroids with intermediate diameters of 30--60 km
(Fig.~\ref{mbo_pole_distribution_diam}c) is also somewhat clustered towards
higher latitudes but the bins for small latitudes are more populated. Therefor, it is
probably a transition region between the two distinct distributions.

\begin{figure}
	\begin{center}
	 \resizebox{\hsize}{!}{\includegraphics{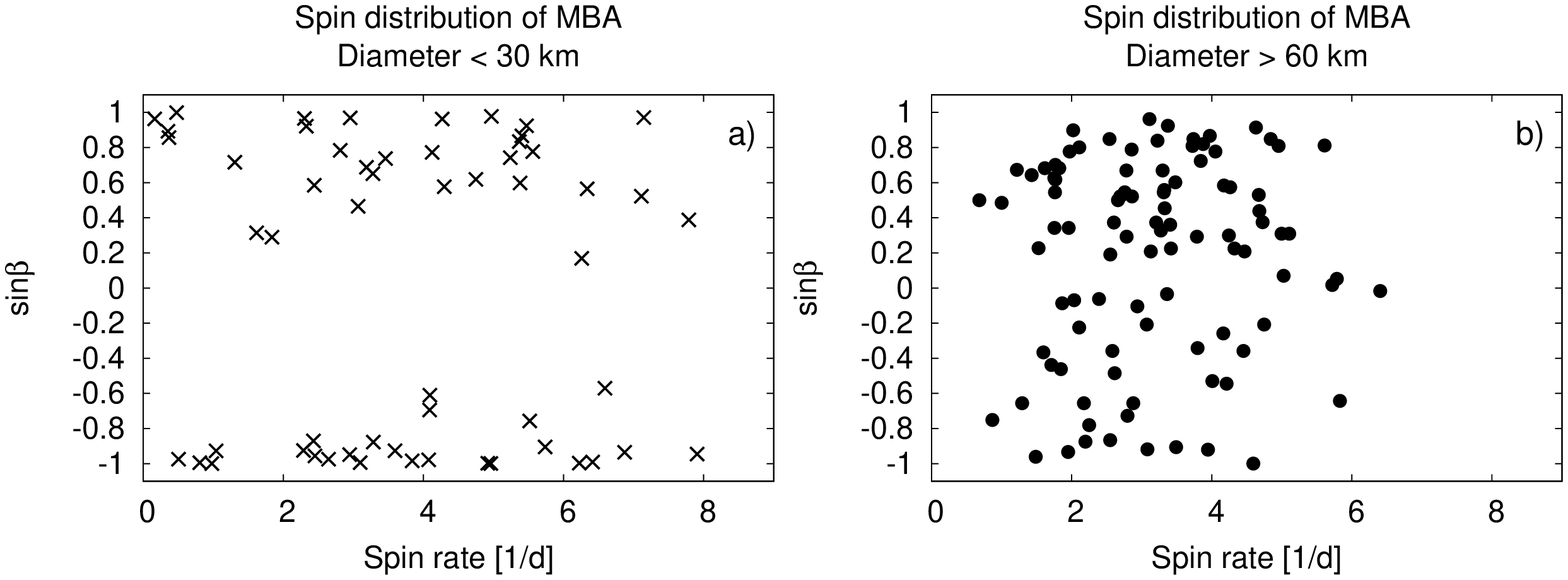}}\\
	\end{center}
	\caption{\label{P_vs_beta}Dependence of the ecliptic latitude $\beta$
(plotted as $\sin\beta$) of the pole direction on the models spin rate, a) for
asteroids with diameter $D < 30$~km and b) for asteroids with diameter $D >
60$~km.}
\end{figure}

It is evident that the depopulation concerns mainly objects with diameters $D
\lesssim 30$~km (the distribution for the intermediate size sample shows that
the limit is probably $\sim$50~km). This size roughly corresponds to the value,
when the YORP effect starts to act and hence it is a natural candidate for a
physical explanation. It is known from previous studies \citep{Pravec2000,
Rubincam2000} that the YORP effect is significantly altering the {\em periods}
and also {\em spin vectors} of these objects on a timescale shorter than the
typical collisional lifetime of these objects (timescales are discussed in more
details in Section~\ref{YORP_sim}). In Fig.~\ref{P_vs_beta}, we show the
relation between the spin rate and the latitude for the small ($D < 30$~km) and
large ($D > 60$~km) groups of asteroids. In concert with YORP theory,
the spin up and spin down and the simultaneous evolution of the latitudes
towards higher absolute values are evident in the small asteroid sample.

Note that observed latitude and longitude distributions can be biased by the
convex inversion method and, therefore, this bias should be taken into
consideration. In general, models for asteroids with higher amplitudes are more
often successfully derived than those asteroids with lower amplitudes. This is because, while the
accuracy of the sparse photometry in both cases is roughly the same, the signal-to-noise ratio is significantly better for higher amplitude lightcurves. If we
assume two bodies with the same shapes and orbits and {\em different} ecliptic
latitudes of the poles, the body with higher absolute value of ecliptic
latitude usually has a higher amplitude. This effect is numerically investigated
in Section \ref{simulations}.

\citet{Skoglov2002} discussed the role of orbital inclination in the observed distribution of latitudes. No indications of this effect were found in the asteroids sample of \citet{Kryszczynska2007}. We also did not find any indications of such
correlation in our sample and so we conclude that orbital inclination does not affect the observed distribution of latitudes.

We are not aware of any other physical effects in the main belt that
could explain the non-uniform observed latitude distribution of small asteroids
($D <$ 30 km). Collisions are believed to produce uniform spin distributions and
close encounters with planets are common only among NEAs.

There are many additional selection effects that influence the properties of
derived models, e.g. the role of amplitude, orbit, time, accuracy and geometry
of observations, among others. The significance of this bias is unknown and cannot
be easily determined. The main problem here is that for almost every asteroid,
the photometric data are from different observers with a different number of
measurements, quality, and purpose. The significance of this effect can be
determined only from a comparison of models derived from real and synthetic data
of known properties. This may be possible in a few years when the photometry
from the Pan-STARRS is available, but not now. In the meantime, the role of the selection effects seems to be small and does not significantly affect, for example, the latitude distribution.

The longitude distributions of the MBAs are plotted in
Figs.~\ref{mbo_pole_distribution}b and ~\ref{mbo_pole_distribution_diam}b,d,f.
They are, contrary to the latitude distributions, without any statistically
significant features and have very close to uniform distributions. The only
exception are the asteroids with $D$~$<$~30 km, but the excess appears to be just a
random coincidence than a result of some physical process.

In all cases, we tested a hypothesis that the observed distribution of latitudes
or longitudes is uniform (using a $\chi^2$--test\footnote{The results based on
the $\chi^2$--test are also in agreement with the Kolmogorov--Smirnov test}).
The computed chi--squares and corresponding probabilities are listed in Table
\ref{chi2}. 
Higher $\chi^2$-values and lower probabilities mean that the supposed hypothesis
``the observed distribution is uniform'' does not fit the observed data. If we
assume a probability of 5\% or lower as statistically significant, we can say
that the latitude distributions for the whole sample and for asteroids with
$D$~$<$~30 km and 30~$<$~$D$~$<$~60~km do not agree with a uniform distribution.
On the other hand, all longitude distributions are consistent with uniform
distributions. Latitude distribution for the MBAs with diameters $D$~$>$~60~km
also disagrees with the uniform distribution; this is because of the excess of
prograde rotators.

The overall view on the model positions within the main belt of asteroids,
together with their estimated total drifts and the information about whether they are
prograde or retrograde rotators (Fig.~\ref{a_vs_e}), shows behavior consistent
with the Yarkovsky/YORP theory: 
there is an asymmetry of prograde and retrograde rotators near the main
resonances and prograde asteroids drift outwards from the Sun and can reach the
resonance only from the left. On the other hand, retrograde rotators drift the
opposite direction and depopulate the zone left of the resonance because the resonance
prevents the entry of new retrograde asteroids. This creates an excess of
prograde rotators. The same mechanism works also in the zone right of the resonance,
except, in this case, only an excess of retrograde asteroids is now created. This effect is obvious
in the neighborhood of the $\nu_6$ and 3:1 resonances.  The total drift from an asteroid's original location during the collisional lifetime computed with
Eq.~\ref{tau_reor}
are inversely proportional to the size of the asteroids. Larger asteroids
($D \gtrsim 50$ km) do not drift significantly while smaller
asteroids frequently drift $\pm$0.05~AU. Note that small-sized asteroids are
found mainly in the inner or middle part of the main belt (due to selection
effect; they have high albedos and/or are closer to the Earth). Asteroids drifting
through the resonances during their collisional lifetime are interesting from
the point of the dynamical evolution. These asteroids were either recently
collisionally-affected or their shape models are wrong. There seems to be three
such models, two near the 3:1 resonance and one near the resonance 5:2.

\begin{table}\caption{\label{chi2}Test of the hypothesis that the observed
pole distributions and de-biased latitude distribution are drown from uniform distributions (a $\chi^2$--test).}
\begin{center}
 \begin{tabular}{crrrrrr}\hline
   & \multicolumn{2}{c} {$\beta$} & \multicolumn{2}{c} {$\lambda$} &
\multicolumn{2}{c} {$\beta_{\mathrm{deb}}$}\\
   & \multicolumn{2}{c} {$N$ = 9} & \multicolumn{2}{c} {$N$ = 11} &
\multicolumn{2}{c} {$N$ = 9}\\ \hline\hline
  Diameter & $\chi^2$ & \% & $\chi^2$ & \% & $\chi^2$ & \% \\
  all & 63 & 10$^{-9}$ & 5.9 & 88 & 39 & 10$^{-3}$ \\ 
  $>60$~km & 17 & 4.5 & 8.2 & 69 & 17 & 4.7 \\ 
  $30-60$~km & 30 & 10$^{-2}$ & 5.6 & 90 & 18 & 2.9 \\ 
  $<30$~km & 59 & 10$^{-7}$ & 20 & 4.8 & 37 & 10$^{-3}$ \\ 
 \end{tabular}
\tablefoot{
$N$ is the degree of freedom.
}
\end{center}
\end{table}

\subsection{The Koronis family members}\label{koronis}

The analysis of rotational state solutions for ten members of the Koronis
asteroid family revealed a clustered distribution of their spin vectors
\citep{Slivan2002,Slivan2003}. This spin distribution was later explained by
\citet{Vokrouhlicky2003} as the result of the thermal torques and spin-orbital
resonances that modify the spin states over time of 2--3~Gyr. The modeling
suggested an existence of two groups of asteroids: (a) low-obliquity retrograde
objects with rotational periods $P<5$~h or $P>13$~h, and (b) prograde rotators with
periods $4<P<7$~h that became trapped in a spin-orbit resonance with secular
frequency s$_6$ and thus have similar spin obliquities ($42-51^{\circ}$) and
also similar ecliptic longitudes in the range of ($24-73^{\circ}$) and
($204-259^{\circ}$). All ten members of the Koronis family studied by
\citet{Slivan2002} and \citet{Slivan2003} had the expected properties: periods
for prograde rotators were shifted only to higher values of 7--10~h.
\citet{Slivan2009} published spin state solutions for another four members of
Koronis family. Only the solution for (253) Dresda was not in agreement with the
theoretical expectation.

Here, we present three new models of asteroids belonging to the Koronis family:
(832) Karin, (1482) Sebastiana, and (1635) Bohrmann, along with two partial models
for (1350) Rosselina and (1389) Onnie. Only the spin state solutions for
Bohrmann and Onnie fit the theoretical expectations. Rotational parameters for
Karin ($P=18.3512$~h, $\lambda=242^{\circ}$, $\beta=46^{\circ}$) are outside
both groups. Asteroids Sebastiana and Rosselina are low-obliquity retrograde
rotators, but their periods (10.49~h for Sebastiana and 8.14~h for Rosselina)
are in the middle of the ``wrong'' range of $P=$ 5--13 hours. Karin is the
namesake and largest member of a small and young \citep[$\sim5.8$ My,
][]{Nesvorny2004} collisional family that is confined within
the larger Koronis family. The spin state of Karin was thus likely affected
during this catastrophic event and changed to a random state that disagrees with
the clustered distribution. 

We are not able to give a satisfactory explanation for the peculiar spin state solutions for Sebastiana and Rosselina. Nevertheless, we are aware of two possible
scenarios: (i) the initial rotational state and shape did not allow being captured
in the resonance or (ii) the objects were randomly reoriented by
non-catastrophic collisions. The timescales of such collisions (given by
Eq.~\ref{tau_reor}) are for Sebastiana $\tau_{\rm reor}\sim 7.5$~Gyr and for
Rosselina $\tau_{\rm reor}\sim 14.7$~Gyr. This leads to the probability of a
collision during the Koronis cluster lifetime \citep[estimated to $\sim$2.5~Gyr,
][]{Bottke2001} $\sim$1/3 for Sebastiana and $\sim$1/6 for Rosselina,
respectively, which means that random collisional reorientation of the spin axis
is likely for at least a few of 19 asteroids in the Koronis cluster with known
spin state solutions (most of them have $\tau_{\rm reor}\lesssim
20$~Gyr).

\subsection{Biases of the LI method}\label{simulations}

\begin{figure}
	\begin{center}
	 \resizebox{\hsize}{!}{\includegraphics{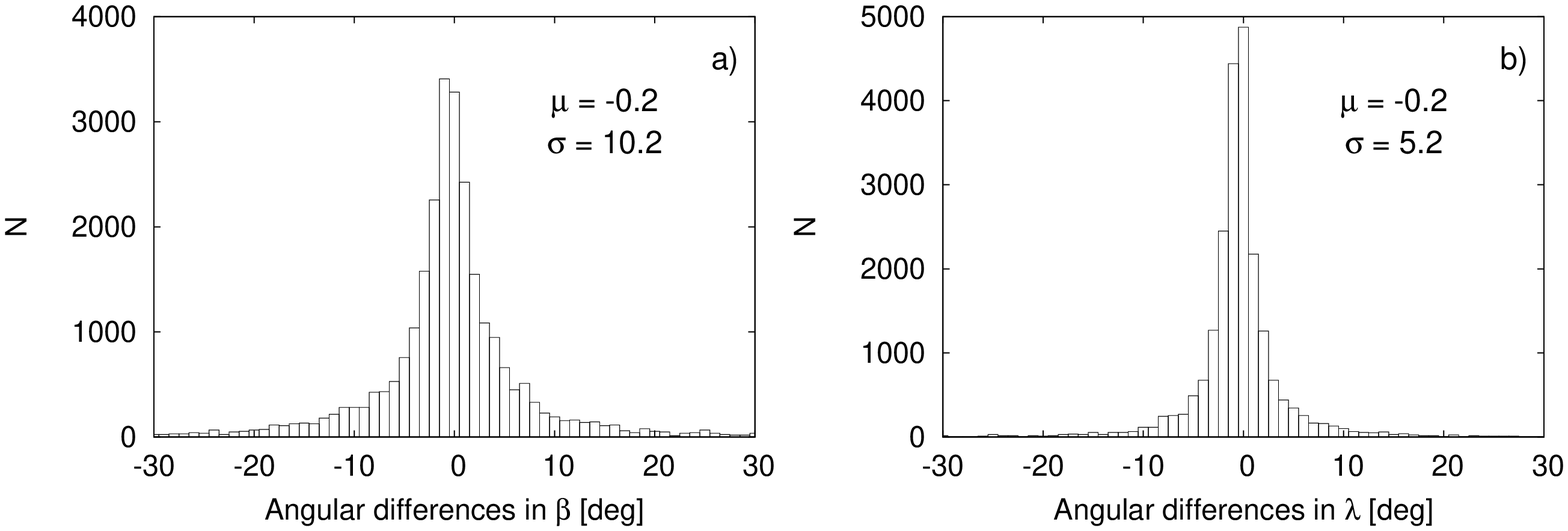}}\\
	\end{center}
	\caption{\label{errors}Histograms of the angular differences between the
generated and derived ecliptic latitudes and longitudes of the pole directions
for all successfully derived synthetic models: a)
$\beta_{\mathrm{comp}}-\beta_{\mathrm{gen}}$, where $\beta_{\mathrm{gen}}$ and
$\beta_{\mathrm{comp}}$ are generated and computed ecliptic latitudes and
similarly b)
$(\lambda_{\mathrm{comp}}-\lambda_{\mathrm{gen}})\cos\beta_{\mathrm{gen}}$ for
ecliptic longitudes.}
\end{figure}

\begin{figure}
	\begin{center}
	 \resizebox{\hsize}{!}{\includegraphics{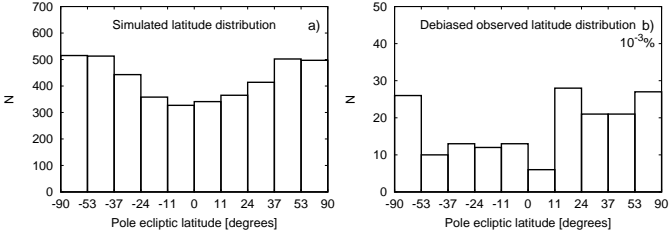}}\\
	\end{center}
	\caption{\label{beta_effect_sim}Fig.~a) shows the simulated ecliptic
latitude distribution of all successfully derived models. In Fig.~b), the
de-biased observed latitude distribution is plotted. The bins are equidistant in
sin$\beta$. In the top right corner, there is the probability value of the
$\chi^2$--test (that the observed distribution is drown from a uniform one, see
Table \ref{chi2}).}
\end{figure}

\begin{figure}
	\begin{center}
	
\resizebox{\hsize}{!}{\includegraphics{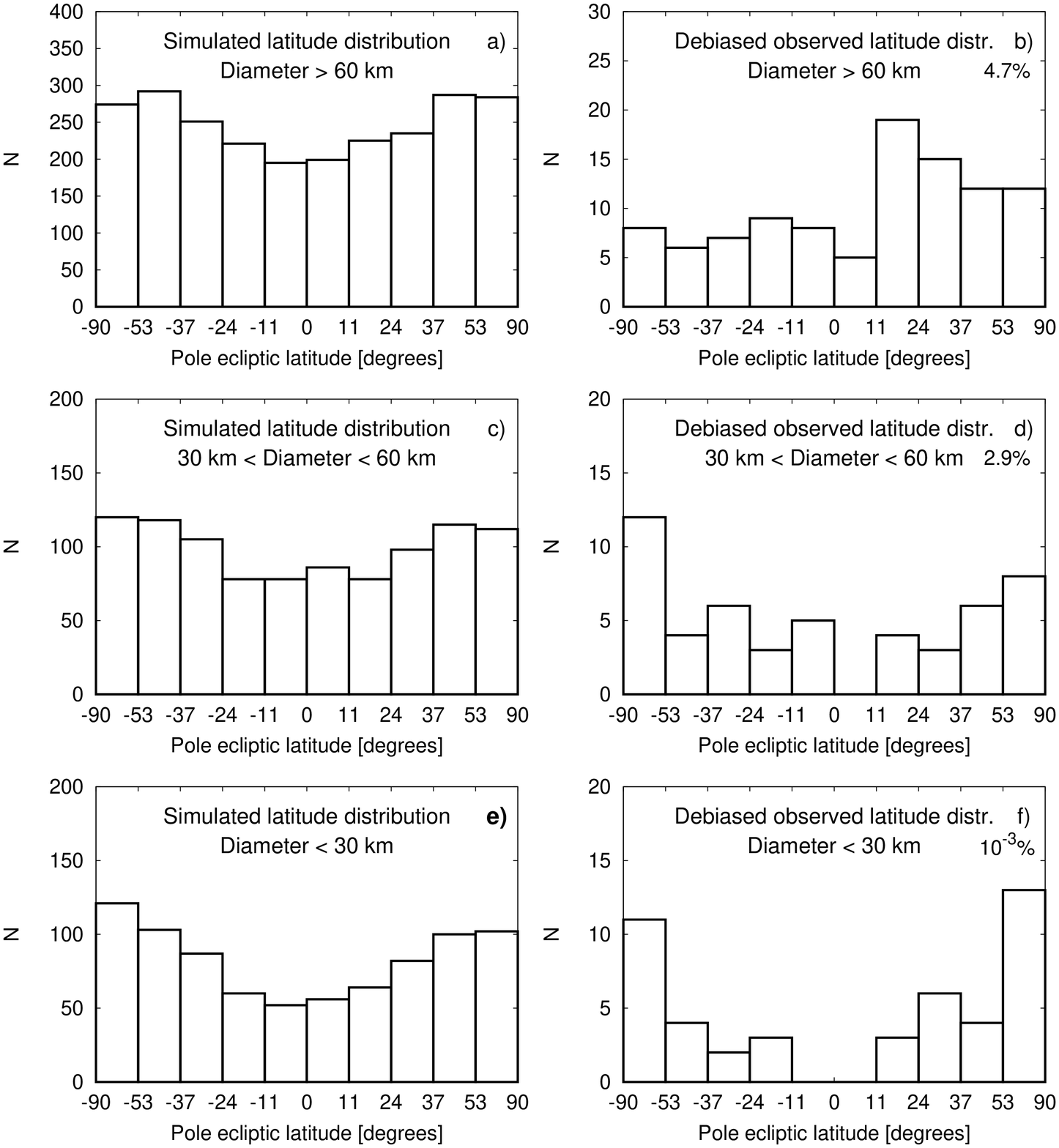}}\\
	\end{center}
	\caption{\label{beta_effect_sim_diam}Histograms showing the simulated
and corrected (de-biased) observed latitude distributions of MBAs for different
size groups, excluding the Koronis cluster members. Fig.~a) shows the simulated
latitude distribution for asteroids with the diameters larger than 60~km, c) for
asteroids with the size in range of 30--60~km, e) for asteroids with diameters
smaller than 30~km. Similarly Figs.~b), d) and f) show the observed latitude
distributions corrected by the bias of the LI method. The bins are equidistant
in sin$\beta$. In the top right corners, there are the probability values of the
$\chi^2$--tests (that the observed distributions are drown from a uniform
distribution, see Table \ref{chi2}).}
\end{figure}

We developed a numerical algorithm to estimate the selection effect of the
lightcurve inversion method and used this approach to de-bias the observed
distribution of asteroid's pole directions. The algorithm was as follows:

\begin{enumerate}
  \item For a model with a known shape, we randomly generated a new pole
direction (while the overall distribution of poles was isotropic).
  \item For each shape with a new rotational state but with the period unchanged, we computed synthetic lightcurves for the same epochs as observed ones.
  \item To each data point $i$, we added the corresponding noise $\delta_i$
given by:
    \begin{center}
    \begin{equation}
    \delta_i=\frac{{L^{\mathrm{obs}}_i}-L^{\mathrm{mod}}_i}{L^{\mathrm{mod}}_i},
    \end{equation}
    \end{center}
  where $L^{\mathrm{obs}}_i$ is $i$-th brightness observed and
$L^{\mathrm{mod}}_i$ is $i$-th brightness computed, both for the original model.
This gave us synthetic lightcurve data equivalent to the original observed data,
but for a new pole direction.
  \item Finally, we performed a lightcurve inversion the same way as with the actual data and tried to derive a model.
  \item We repeated steps 1--4 for 50 random poles for each asteroid model.
\end{enumerate}

In this simulation, we used 80 models derived from combined dense and sparse
data sets and 89 models from the DAMIT.

For each successfully derived model we have the generated pole direction
($\lambda_{\mathrm{gen}}$, $\beta_{\mathrm{gen}}$) and period
$P_{\mathrm{gen}}$, and also the computed pole and period:
$\lambda_{\mathrm{comp}}$, $\beta_{\mathrm{comp}}$,  $P_{\mathrm{comp}}$. We
computed the angular differences between the generated and derived ecliptic
latitude and longitude of the pole direction:
$\beta_{\mathrm{comp}}-\beta_{\mathrm{gen}}$ and
$(\lambda_{\mathrm{comp}}-\lambda_{\mathrm{gen}})\cos\beta_{\mathrm{gen}}$ (the
$\cos\beta_{\mathrm{gen}}$ factor is used for the correction of the different
distances of meridians near the equator and poles). In Fig.~\ref{errors}, we
show the histograms of these differences in a) ecliptic latitudes and b)
ecliptic longitudes. We assumed that the histograms can be described by a normal
distribution and we computed the mean and the standard deviation ($\mu$;
$\sigma$). We found values of ($-$0.2;10.2) for latitudes and ($-$0.2;5.2) for
longitudes. The standard deviation $\sigma$ is directly related to the typical
uncertainty that we can expect in pole determination by the lightcurve inversion
method, which is $\sim$5$^{\circ}/\cos\beta$ in $\lambda$ and
$\sim$10$^{\circ}$ in $\beta$.

In Fig.~\ref{beta_effect_sim}a, we constructed a histogram of the latitude
distribution for all successful models. The bins in $\beta$ were again equally
spaced in $\sin\beta$. The latitude distribution of all generated models was not
exactly uniform, the amount of latitudes in bins slightly differed. To remove
this effect, we divided the latitude distribution of successfully derived models
by the latitude distribution of all generated models normalized to unity. This
correction was also applied to latitude distributions in
Figs.~\ref{beta_effect_sim_diam}a,c,e. It is obvious that the LI method is more
efficient for asteroids with higher $|\beta|$. The amount of successfully
derived models with $|\beta|\sim $0$^{\circ}$ is about 30\% lower than with
$|\beta| >$ 53$^{\circ}$.

In Figs.~\ref{beta_effect_sim_diam}a,c,e, we constructed the histograms of
latitude distributions for successfully derived models and distinguished
three size ranges. All three plots look very similar, except that with
decreasing size, the ratio between models with $|\beta|\sim$0$^{\circ}$ and
$|\beta| >$ 53$^{\circ}$ goes down. This ratio is $\sim$75\% for $D >
60$~km, $\sim$65\% for $30 < D < 60$~km and $\sim$60\% for $D < 30$~km.

The histograms in Figs.~\ref{beta_effect_sim}a and
\ref{beta_effect_sim_diam}a,c,e define the bias in latitude of the LI
method and can be used for de-biasing the observed latitude distributions
presented in Figs.~\ref{mbo_pole_distribution}a and
\ref{mbo_pole_distribution_diam}a,c,e. The de-biased histograms of latitudes are
plotted in Figs.~\ref{beta_effect_sim}b and
\ref{beta_effect_sim_diam}b,d,f. The histograms changed only slightly and the
conclusions from Section~\ref{spin_vector} are still valid, i.e., the latitude
distribution differs significantly from a uniform distribution, and especially so for $D < 30$~km. The distribution of latitudes for asteroids
with $D > 60$ has an evident excess of prograde rotators while the distribution for
a subsample with 30 km $< D <$ 60 km shows an enrichment of
asteroids with large latitudes ($|\beta| >$ 53$^{\circ}$). Other bins have
similar populations. We also performed a $\chi^2$--test in the
same way as for the observed distributions (see Table~\ref{chi2}, columns 6 and
7).

We did not find any significant correlation between the ecliptic longitude and
the efficiency of the model determination.

\section{A theoretical model of the latitude distribution}\label{YORP_sim}

\begin{figure*}
\begin{center}
\includegraphics[width=8cm]{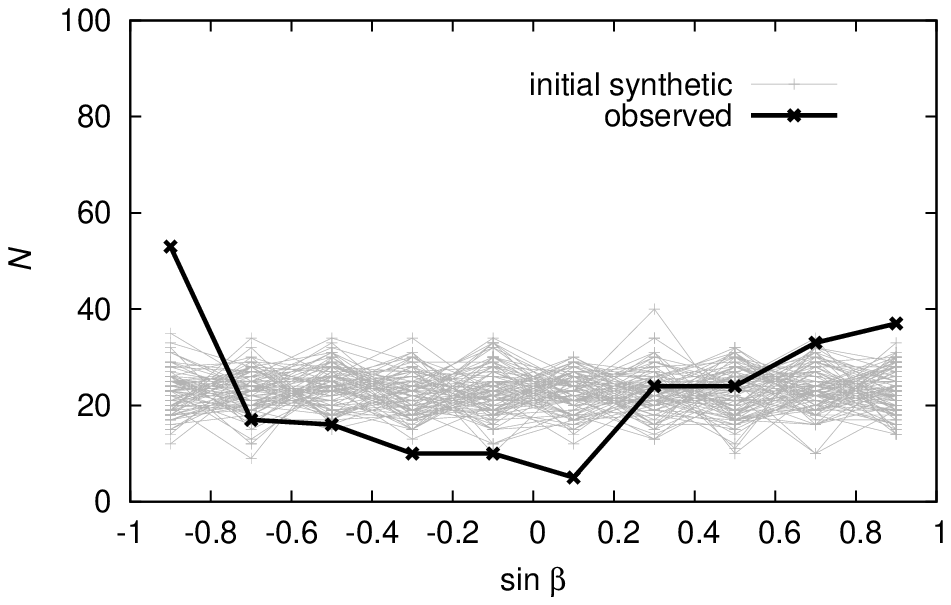}
\includegraphics[width=8cm]{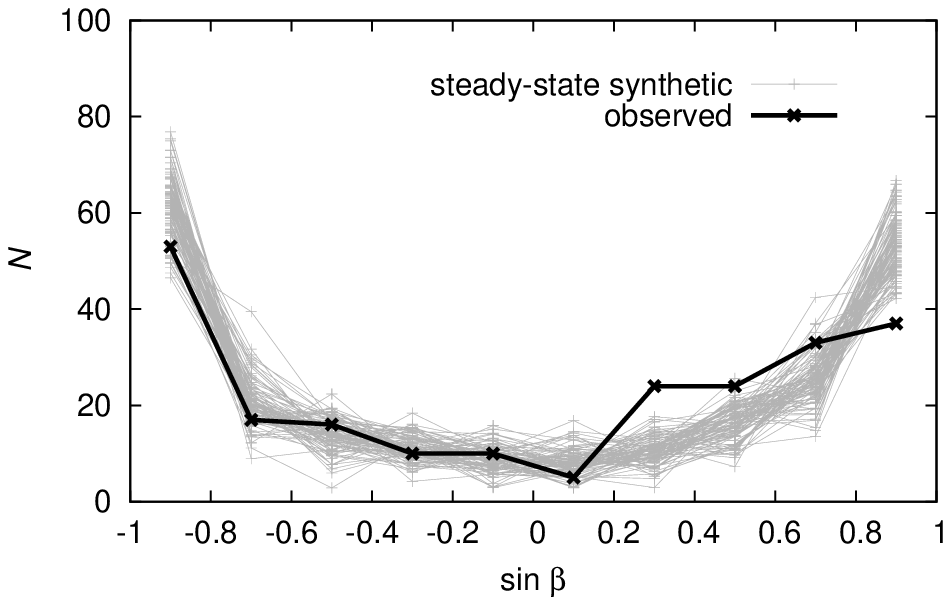}
\end{center}
\caption{Left panel: The distribution of the ecliptic latitudes~$\beta$
for the observed asteroids (thick line) and 100 synthetic samples
(generated with a different random seed) at time $t = 0$ (thin lines).
Right panel: steady-state synthetic latitude distributions at time $t = 4\,{\rm
Gyr}$ (thin lines),
evolved by the YORP effect, collisions, mass shedding and spin-orbit resonances.
We also applied an observational bias to it.
The steady state was reached already within $\simeq 1\,{\rm Gyr}$.}
\label{hanus_ISEED_cYORP_0.33_betahist_resonances}
\end{figure*}

In order to understand observations of main-belt asteroids,
namely the de-biased distribution of their ecliptic latitudes~$\beta$
(Fig.~\ref{beta_effect_sim}b), we constructed a simple model
for spin evolution that contains the following processes:
  (i)~the YORP effect, i.e., torques arising from the emission of thermal
radiation,
 (ii)~random reorientations induced by non-catastrophic collisions, and
(iii)~mass shedding after a critical rotational frequency is reached.

On the other hand, we did {\em not\/} include
gravitational torques of the Sun,
spin-orbital resonances,
damping (dissipation of rotational energy), or tumbling.
Even though individual asteroids may be substantially affected
by these processes, our model is for a large statistical sample of
asteroids and the effect on the {\em overall} latitude distribution
is assumed to be only minor. For example, gravitational
torques and spin-orbital resonances usually cause large oscillations
of~$\beta$ for prograde-rotating asteroids, but they remain bound
to a certain interval \citep{Vokrouhlicky2006}.
Moreover, we tried to account for these (rather random) oscillations
in our model as well (see below).

Our sample of 220 asteroids was the same as the observed sample
discussed in Section~\ref{statistics}. This means that the orbits and sizes
correspond to real asteroids.
The model for spin evolution was similar to that of \citet{Broz2011s},
where it was used for studies of the long-term evolution of asteroid families.
We assumed the following relations for the rate of the angular velocity~$\omega$
and the obliquity~$\epsilon$ due to the YORP effect
\begin{eqnarray}
{{\rm d}\omega\over{\rm d} t} &=& f_i(\epsilon)\,,\qquad i = 1 \dots
200\,,\label{eq:domega}\\
{{\rm d}\epsilon\over{\rm d} t} &=&
{g_i(\epsilon)\over\omega}\,,\label{eq:depsil}
\end{eqnarray}
where $f$- and $g$-functions were given by \citet{Capek2004} for a set of 200
shapes with
mean radius $R_0 = 1\,{\rm km}$,
bulk density $\rho_0 = 2500\,{\rm kg}/{\rm m}^3$,
located on a circular orbit with semi-major axis $a_0 = 2.5\,{\rm AU}$.
We assigned one of the artificial shapes (denoted by the index~$i$) randomly to
each individual asteroid.%
\footnote{We did not use the convex-hull shapes derived in this work for two
reasons:
 (i)~the two samples of shapes are believed to be statistically equivalent
     and it is thus not necessary to compute the YORP torques again;
(ii)~the YORP effect seems sensitive to small-scale surface structure
\citep{Scheeres2007}
     which cannot be caught by our shape model. Nevertheless, the YORP torque
     remains of the same order, so the random assignment of shapes seems
reasonable.}
We had only to scale the $f$- and $g$-functions by a factor
\begin{equation}
c = c_{\rm YORP} \left({a\over a_0}\right)^{-2} \left(R\over R_0\right)^{-2}
\left(\rho_{\rm bulk}\over\rho_0\right)^{-1}\,,
\end{equation}
where $a$, $R$, $\rho_{\rm bulk}$ are semi-major axis, radius, and density of the
simulated body, respectively,
and $c_{\rm YORP}$ is a free scaling parameter, which can account for an
additional uncertainty of the YORP model.
Because the values of $f$'s and $g$'s were computed for only a limited set of
obliquities (with a~step $\Delta\epsilon = 30^\circ$)
we used interpolation by Hermite polynomials \citep{Hill1982} of the data in
\citet{Capek2004}
to obtain a smooth analytical functions for $f_i(\epsilon)$ and $g_i(\epsilon)$.

When the angular velocity approached a critical value (i.e., the gravity was
equal to the centrifugal force)
\begin{equation}
\omega_{\rm crit} = \sqrt{{4\over 3} \pi G \rho_{\rm bulk}}\,,
\end{equation}
we assumed a mass shedding event. We kept the orientation of
the spin axis and the sense of rotation but reset the orbital period~$P =
{2\pi/\omega}$
to a random value from the interval $(P_1, P_2)=(2.5, 9)$~hours.
We also altered the assigned shape since any change of shape can produce a different YORP effect.
We did not change the mass, however.

The differential equations~(\ref{eq:domega}) and (\ref{eq:depsil})
were integrated numerically by a simple Euler integrator.
The usual time step was $\Delta t = 1000\,{\rm yr}$.
The time scale of the spin axis evolution for small bodies ($D \simeq 10\,{\rm
km}$)
is $\tau_{\rm YORP} \simeq 500\,{\rm Myr}$.
After $\simeq3$ times $\tau_{\rm YORP}$ most of these bodies have spin axes
perpendicular to the ecliptic.

We also included a Monte-Carlo model for spin axis reorientations
caused by collisions.%
\footnote{Collisional disruptions are not important, since we are only
interested in the steady state. We can imagine that whenever an asteroid
from our sample is disrupted, another one with a randomly oriented spin axis
is created by a disruption of a larger body.}
We used an estimate of the time scale by \citet{Farinella1998}
\begin{equation}\label{tau_reor}
\tau_{\rm reor} = B \left({\omega\over\omega_0}\right)^{\beta_1} \left({D \over
D_0}\right)^{\beta_2}\,,
\end{equation}
where
$B = 84.5\,{\rm kyr}$,
$\beta_1 = 5/6$,
$\beta_2 = 4/3$,
$D_0 = 2\,{\rm m}$ and
$\omega_0$ corresponds to period $P = 5$~hours.
These values are characteristic for the main belt.
After a collision, we reset the spin axis periods to random values, using the interval $(P'_1, P'_2)=(2.5,
9)$~hours for the period.
Since the time scale is $\tau_{\rm reor} \simeq 3\,{\rm Gyr}$
for the smallest ($D \simeq 5\,{\rm km}$) bodies,
reorientations are only of minor importance.
However, note that the probability of the reorientation is enhanced
when the YORP effect drives the angular velocity~$\omega$ close to zero.

There were several free parameters in our model:
the $c_{\rm YORP}$ parameter,
thermal conductivity~$K$,
bulk density $\rho_{\rm bulk}$,
initial distribution of $\beta$
and initial distribution of $\omega$.

Our aim was to start with a simple $\beta$- and $\omega$-distribution,
wait until a steady state was reached,
and then compare the resulting synthetic to observed latitude distributions.
We applied an observational bias derived in Section~\ref{simulations} to the
synthetic distribution.

We partly accounted for spin-orbital resonances acting on prograde
asteroids by adding a sinusoidal oscillations to~$\beta$ with a random
phase
and an amplitude $\simeq 40^\circ$, which are typically induced by resonances.
This procedure naturally decreased the right-most bin ($\sin\beta = (0.8, 1)$)
of the synthetic distribution and increased the next bin ($\sin\beta = (0.6,
0.8)$).

\begin{figure}
\centering
\includegraphics[width=6cm]{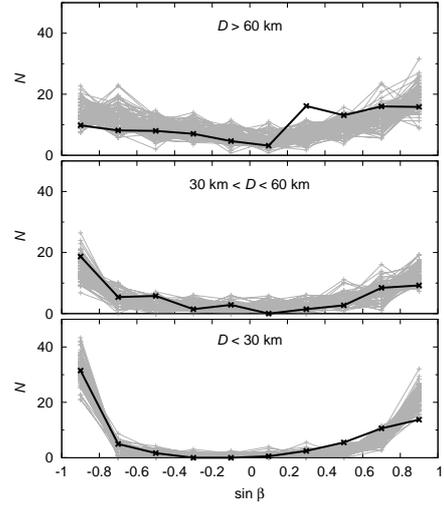}
\caption{Steady-state synthetic latitude distributions for three different size
ranges:
$D > 60\,{\rm km}$ (top),
$30\,{\rm km} < D < 60\,{\rm km}$ (middle) and
$D < 30\,{\rm km}$ (bottom).
The synthetic distributions are plotted by thin gray lines
while the observed distribution by a thick black line.}
\label{hanus_ISEED_cYORP_0.33_betahist_resonances_SIZES}
\end{figure}

We started with reasonable parameters of
$c_{\rm YORP} = 0.33$,
$K = 10^{-2}\,{\rm W}/{\rm K}/{\rm m}$,
$\rho_{\rm bulk} = 2500\,{\rm kg}/{\rm m}^3$,
a Maxwellian distribution of~$\omega$,
a uniform distribution of~$\sin\beta$ (i.e., an isotropic distribution of
spin axes).
We ran 100 such simulations with different random seeds.
A steady state was reached within $\simeq 1\,{\rm Gyr}$.
The resulting latitude distributions are shown in
Figures~\ref{hanus_ISEED_cYORP_0.33_betahist_resonances}
and~\ref{hanus_ISEED_cYORP_0.33_betahist_resonances_SIZES}.

From these it can be seen that:
(i)~the observed distribution of $\beta$ for small asteroids seems compatible
with our model; the YORP effect is capable of creating such an uneven distribution and 
(ii)~there is a discrepancy for large asteroids (especially in bins $\sin\beta
\in (-1, -0.8)$ and $(0.2, 0.4)$),
which can be explained as a preference for prograde rotators
in the primordial population \citep[see ][]{Davis1989,
Johansen2010}.
The results regarding the spin rates agree with \citet{Pravec2008},
so we do not repeat the discussion here.

We also tested the sensitivity of our results with respect to the free parameters.
The thermal conductivity did not seem important (we tested $K = 10^{-3}\,{\rm
W}/{\rm K}/{\rm m}$).
A simulation with
$c_{\rm YORP} = 0.66$,
$\rho = 1300\,{\rm kg}/{\rm m^3}$,
and a uniform distribution of orbital periods $P \propto 1/\omega$
produced almost the same resulting latitude distribution.
Nevertheless, a value of $c_{\rm YORP} = 1.00$ seems too high because
the extreme bins of the $\beta$-distribution were
overpopulated.\footnote{As an alternative hypothesis, we assumed the spin
axis evolution without a YORP effect ($c_{\rm YORP} = 0$). In this case, the
initial $\beta$-distributions (Fig.
~\ref{hanus_ISEED_cYORP_0.33_betahist_resonances}, left panel, thin lines) do
not change significantly in time.} 
There is only a weak dependence of our results on the period ranges that we used for resetting the orbital period after a
mass-shedding event ($P_1, P_2$) and collision ($P'_1, P'_2$). As would be expected, the values $P_1, P_2$, $P'_1,
P'_2$ significantly affect the period distribution.
The relatively weak dependence on the free parameters likely
stems from the fact that we presume the steady-state.
Even though the free parameters change, e.g., the strength
of the YORP effect and the evolution of spins is slower and/or faster,
after reaching a steady state, the basic characteristics of the
latitude distribution remain similar.
The observed $\beta$-distribution of small asteroids ($D <$ 30 km)
cannot
be explained by our simulation without accounting the YORP effect.

\section{Conclusions}

The results of this paper can be summarized as follows.

We used combined dense and sparse data to derive new asteroid shape models. We
systematically gathered and processed all available sparse photometry from
astrometric surveys and employed valuable data from seven observatories (see
Table \ref{observatories}) in lightcurve inversion.

We derived 80 new unique models of asteroids, from which 16 are based only on
sparse data. We also present 30 partial models with accurate rotational periods
and estimated ecliptic latitudes of the pole directions and 18 updated solutions
based on new data for asteroids already included in DAMIT.

In the future, quality sparse data sets will be produced by all-sky
surveys such as Pan-STARRS, the Large Synoptic Survey Telescope (LSST), and the Gaia satellite. When these data are available, we will be able to apply the same methods to in order to derive many more new asteroid models. 
These surveys will have one advantage over dense data: their selection
effects (e.g., with respect to the orbit) will be known. This will allow us to
make a more accurate analysis of the asteroid population.

As expected, the observed ecliptic {\em longitude} distribution of asteroid spin
vector is independent of diameter and is compatible with a
uniform distribution. Unlike the latitude distribution, the observed ecliptic longitude distribution is not significantly biased by the LI method. However,
the effect of the LI bias is only minor and the global features of the
observed latitude distribution do not change. The observed (and de-biased) {\em latitude} distribution for asteroids with diameters $D >60$~km shows an
excess of prograde rotators in the latitude interval (11$^\circ$,
90$^\circ$). This excess is probably primordial. On the other hand, the latitude
distributions for the entire sample and in particular for asteroids with $D <$ 30 km, is strongly anisotropic.

The dynamical evolution of asteroid spins seems to be dominated by the YORP
effect and also by collisions and mass shedding for asteroids with diameters $D
\lesssim$ 30 km. We calculate that YORP (with a small contribution for the LI
method's bias) is capable of producing the observed depopulation of spin vectors
for small asteroids.

We are not yet able to study small asteroids in individual families (small
bodies at the outskirts of a family should have extreme spins); this is an aim
of future work.

\begin{acknowledgements}
The work of JH has been supported by the grant GA\,UK 134710 of the Grant
agency of the Charles University, by the project SVV 261301 of the Charles
University in Prague and by the grant GACR 205/08/H005 of the Czech grant
agency. The work of JH and J\v D has been supported by the grant GACR
209/10/0537 of the Czech grant agency, the work of JD and MB by the Research
Program MSM0021620860 of the Czech Ministry of Education and the work of MB has
been also supported by the Grant Agency of the Czech Republic (grant
205/08/P196). The calculations were performed on the computational cluster Tiger
at the Astronomical Institute of Charles University in Prague
(\texttt{http://sirrah.troja.mff.cuni.cz/tiger}).

We thank Brian D.~Warner for significantly improving the linguistic quality of this paper and David \v Capek for sending us the YORP effect data in an electronic form.
\end{acknowledgements}

\bibliography{mybib}
\bibliographystyle{aa}

\onecolumn

\longtab{3}{
\begin{longtable}{r@{\,\,\,}l rrrr D{.}{.}{6} rrrrrrrrrrc}
\caption{\label{models_tab}List of new asteroid models derived from combined data sets or sparse data alone.}\\
\hline 
\multicolumn{2}{c} {Asteroid} & \multicolumn{1}{c} {$\lambda_1$} & \multicolumn{1}{c} {$\beta_1$} & \multicolumn{1}{c} {$\lambda_2$} & \multicolumn{1}{c} {$\beta_2$} & \multicolumn{1}{c} {$P$} & $N_{\mathrm{lc}}$ & $N_{\mathrm{app}}$  & $N_{\mathrm{689}}$ & $N_{\mathrm{699}}$ & $N_{\mathrm{703}}$ & $N_{\mathrm{E12}}$ & $N_{\mathrm{G96}}$ & $N_{\mathrm{950}}$ & $N_{\mathrm{Hip}}$\\
\multicolumn{2}{l} { } & [deg] & [deg] & [deg] & [deg] & \multicolumn{1}{c} {[hours]} &  &  &  &  &  &  &  &  &\\
\hline\hline

\endfirsthead
\caption{continued.}\\

\hline
\multicolumn{2}{c} {Asteroid} & \multicolumn{1}{c} {$\lambda_1$} & \multicolumn{1}{c} {$\beta_1$} & \multicolumn{1}{c} {$\lambda_2$} & \multicolumn{1}{c} {$\beta_2$} & \multicolumn{1}{c} {$P$} & $N_{\mathrm{lc}}$ & $N_{\mathrm{app}}$  & $N_{\mathrm{689}}$ & $N_{\mathrm{699}}$ & $N_{\mathrm{703}}$ & $N_{\mathrm{E12}}$ & $N_{\mathrm{G96}}$ & $N_{\mathrm{950}}$ & $N_{\mathrm{Hip}}$\\
\multicolumn{2}{l} { } & [deg] & [deg] & [deg] & [deg] & \multicolumn{1}{c} {[hours]} &  &  &  &  &  &  &  &  &\\
\hline\hline
\endhead
\hline
\endfoot
10 & Hygiea & 312 & $-$42 & \textbf{122} & \textbf{$-$44} & 27.6591 & 23 & 9 & 263 &   &   &   &   & 405 & 50\\
13 & Egeria & 44 & 21 & 238 & 11 & 7.04667 & 13 & 4 & 255 &   & 74 &   &   & 203 & 34\\
14 & Irene & 97 & $-$22 & 268 & $-$24 & 15.02991 & 20 & 8 & 250 &   &   & 48 &   & 161 & 45\\
37 & Fides & 270 & 19 & 89 & 27 & 7.33253 & 23 & 5 & 270 &   & 61 &   &   & 135 & 31\\
40 & Harmonia & 22 & 31 & 206 & 39 & 8.90848 & 19 & 6 & 210 &   & 48 &   &   & 255 & 102\\
42 & Isis & 106 & 40 & 302 & 28 & 13.58364 & 28 & 7 & 210 &   & 36 &   &   & 128 & 51\\
62 & Erato & 87 & 22 & 269 & 23 & 9.21819 & 1 & 1 & 164 &   & 48 &   &   &   & \\
68 & Leto & 103 & 43 & 290 & 23 & 14.84547 & 12 & 2 & 174 &   & 85 & 30 &   & 152 & \\
69 & Hesperia & 250 & 17 & 71 & $-$2 & 5.65534 & 35 & 7 & 222 &   & 44 & 40 &   &   & \\
97 & Klotho & $-$1 & 30 & 161 & 40 & 35.2510 & 25 & 6 & 309 &   & 31 &   &   & 202 & \\
119 & Althaea & 339 & $-$67 & 181 & $-$61 & 11.46514 & 4 & 2 & 149 &   & 59 &   &   & 222 & \\
162 & Laurentia & 139 & 64 & 313 & 51 & 11.86917 & 4 & 2 & 166 & 31 & 40 &   &   &   & \\
174 & Phaedra & 94 & 36 & 266 & 14 & 5.75025 & 2 & 1 & 173 &   & 36 &   &   &   & \\
188 & Menippe & 32 & 48 & 198 & 25 & 11.9765 & 4 & 1 & 145 &   & 40 &   &   &   & \\
258 & Tyche & 224 & $-$4 & 40 & $-$9 & 10.04008 & 10 & 2 & 162 &   & 44 &   &   &   & \\
264 & Libussa & 157 & 18 & 338 & $-$9 & 9.22794 & 19 & 3 & 129 &   & 39 & 49 &   &   & \\
291 & Alice & 69 & 51 & 249 & 56 & 4.316011 & 9 & 4 & 75 &   & 46 &   &   &   & \\
302 & Clarissa & \textbf{28} & \textbf{$-$72} & 190 & $-$72 & 14.47670 & 8 & 2 & 102 &   & 104 &   &   &   & \\
310 & Margarita & 225 & $-$35 & 42 & $-$33 & 12.0710 & 27 & 1 & 88 & 31 & 51 &   &   &   & \\
312 & Pierretta & 82 & $-$39 & 256 & $-$58 & 10.20764 & 4 & 1 & 176 &   & 36 & 52 &   &   & \\
336 & Lacadiera & 194 & 39 & 37 & 54 & 13.69552 & 3 & 1 & 121 &   & 36 & 32 &   &   & \\
340 & Eduarda & 188 & $-$43 & 18 & $-$47 & 8.00613 & 2 & 1 & 117 &   & 76 & 31 & 36 &   & \\
354 & Eleonora & 144 & 54 &   &   & 4.277186 & 37 & 9 & 258 &   & 40 &   &   & 139 & 96\\
355 & Gabriella & 341 & 78 & 197 & 70 & 4.82899 & 4 & 1 & 128 &   &   &   &   &   & \\
367 & Amicitia & 203 & 38 & 21 & 32 & 5.05502 & 2 & 1 & 128 &   & 34 &   &   &   & \\
372 & Palma & 221 & $-$47 & 44 & 17 & 8.58189 & 28 & 6 & 214 &   & 52 & 36 &   &   & \\
376 & Geometria & 239 & 45 & 63 & 53 & 7.71097 & 39 & 9 & 158 &   & 76 &   &   &   & \\
399 & Persephone & 36 & 63 &  &  & 9.14639 &   &   & 166 &   & 36 &   &   &   & \\
400 & Ducrosa & 328 & 56 & 158 & 62 & 6.86788 & 3 & 1 & 103 &   &   &   &   &   & \\
413 & Edburga & 202 & $-$45 &   &   & 15.7715 & 2 & 1 & 148 &   & 43 &   &   &   & \\
436 & Patricia & 124 & $-$30 & 339 & $-$58 & 16.1320 & 4 & 1 & 97 &   & 53 & 91 &   &   & \\
440 & Theodora & 80 & $-$88 &  &  & 4.83658 &   &   & 123 &   & 103 &   & 48 &   & \\
471 & Papagena & \textbf{223} & \textbf{67} & 22 & 18 & 7.11539 & 13 & 2 & 293 &   & 72 &   &   & 203 & 112\\
486 & Cremona & 227 & 59 & 31 & 30 & 65.151 & 1 & 1 & 127 &   & 55 & 35 &   &   & \\
499 & Venusia & 37 & 50 & 212 & 46 & 13.4871 & 4 & 1 & 122 &   & 39 &   & 31 &   & \\
544 & Jetta & 275 & $-$84 & 31 & $-$67 & 7.74528 & 3 & 1 & 139 &   & 60 &   &   &   & \\
573 & Recha & 74 & $-$24 & 252 & $-$48 & 7.16586 & 3 & 1 & 161 &   & 85 &   &   &   & \\
584 & Semiramis & 106 & $-$56 & 315 & $-$32 & 5.06893 & 24 & 6 & 150 &   & 59 & 49 &   &   & \\
590 & Tomyris & 273 & $-$47 & 120 & $-$46 & 5.55248 & 3 & 1 & 91 &   & 32 &   &   &   & \\
601 & Nerthus & 173 & 44 & 20 & 32 & 13.5899 &   &   & 139 &   & 94 &   &   &   & \\
606 & Brangane & 183 & 20 & 354 & 26 & 12.29067 & 2 & 1 & 108 &   & 70 &   &   &   & \\
629 & Bernardina & 40 & 33 & 236 & 48 & 3.76360 &   &   & 91 &   & 48 &   &   &   & \\
631 & Philippina & 183 & $-$2 &   &   & 5.90220 & 6 & 2 & 171 &   & 38 &   &   &   & \\
685 & Hermia & 197 & 87 & 29 & 79 & 50.387 &   &   & 93 &   & 148 &   &   &   & \\
695 & Bella & 87 & $-$55 & 314 & $-$56 & 14.21900 & 8 & 1 & 184 &   & 90 & 30 &   &   & \\
753 & Tiflis & 5 & 36 & 199 & 57 & 9.8259 &   &   & 129 &   & 64 &   &   &   & \\
800 & Kressmannia & 345 & 37 & 172 & 34 & 4.460963 & 8 & 2 & 108 &   & 51 &   &   &   & \\
808 & Merxia & 26 & 54 & 192 & 57 & 30.630 & 4 & 1 & 158 &   & 87 &   & 32 &   & \\
810 & Atossa & 12 & 67 & 188 & 69 & 4.38547 &   &   & 99 &   & 71 &   & 60 &   & \\
825 & Tanina & 46 & 48 & 231 & 60 & 6.93981 & 2 & 1 & 114 &   & 40 &   &   &   & \\
832 & Karin & 242 & 46 & 59 & 44 & 18.3512 & 13 & 3 & 84 &   &   & 39 &   &   & \\
847 & Agnia & 341 & 18 & 162 & 13 & 14.8247 & 3 & 1 & 136 &   &   &   &   &   & \\
889 & Erynia & 187 & $-$60 & 335 & $-$74 & 9.8749 &   &   & 94 &   & 65 &   &   &   & \\
925 & Alphonsina & \textbf{296} & \textbf{41} & 147 & 22 & 7.87754 & 4 & 1 & 134 &   & 48 & 79 &   &   & \\
934 & Thuringia & 120 & $-$52 &   &   & 8.16534 &   &   & 123 &   & 59 &   &   &   & \\
1002 & Olbersia & 220 & 35 & 16 & 54 & 10.2367 &   &   & 87 &   & 48 & 54 &   &   & \\
1087 & Arabis & 334 & $-$7 & 155 & 12 & 5.79501 & 3 & 1 & 156 &   & 92 &   &   &   & \\
1102 & Pepita & 25 & $-$34 & 231 & $-$30 & 5.10532 &   &   & 147 &   & 47 &   &   &   & \\
1140 & Crimea & 12 & $-$73 & 175 & $-$22 & 9.7869 & 3 & 1 & 96 &   & 116 &   &   &   & \\
1148 & Rarahu & 148 & $-$9 & 322 & $-$9 & 6.54449 &   &   & 95 &   & 64 &   &   &   & \\
1207 & Ostenia & 310 & $-$77 & 124 & $-$51 & 9.07129 & 2 & 2 & 87 &   & 71 &   &   &   & \\
1291 & Phryne & 106 & 35 & 277 & 59 & 5.58414 & 2 & 1 & 129 &   & 72 &   &   &   & \\
1301 & Yvonne & 39 & 41 &   &   & 7.31968 &   &   & 78 &   & 56 & 33 &   &   & \\
1333 & Cevenola & 38 & $-$86 & 220 & $-$44 & 4.87932 & 3 & 1 & 104 &   & 91 &   &   &   & \\
1382 & Gerti & 268 & 23 & 87 & 28 & 3.081545 & 2 & 1 & 60 &   & 56 &   & 52 &   & \\
1419 & Danzig & 22 & 76 & 193 & 62 & 8.11957 & 1 & 1 & 135 &   & 87 &   &   &   & \\
1482 & Sebastiana & 262 & $-$68 & 91 & $-$67 & 10.48965 & 2 & 1 & 131 &   & 39 & 30 &   &   & \\
1514 & Ricouxa & 251 & 75 & 68 & 69 & 10.42468 & 3 & 1 & 68 &   & 56 &   &   &   & \\
1568 & Aisleen & 109 & $-$68 &   &   & 6.67597 &   &   & 82 &   & 37 &   &   &   & \\
1635 & Bohrmann & 5 & $-$38 & 185 & $-$36 & 5.86427 & 8 & 1 & 108 &   & 47 &   &   &   & \\
1659 & Punkaharju & 259 & $-$71 & 75 & $-$22 & 5.01327 & 2 & 1 & 118 &   & 66 &   &   &   & \\
1682 & Karel & 232 & 32 & 51 & 41 & 3.37485 &   &   & 54 &   & 84 &   & 36 &   & \\
1709 & Ukraina & 165 & $-$61 & 2 & $-$40 & 7.30517 & 2 & 1 & 46 &   & 79 &   &   &   & \\
1742 & Schaifers & 198 & 57 & 47 & 55 & 8.53270 & 3 & 1 & 106 &   &  &   &   &   & \\
1747 & Wright & 227 & 31 &   &   & 5.28796 &   &   & 70 &   & 55 &   &   &   & \\
1889 & Pakhmutova & 22 & $-$76 & 167 & $-$40 & 17.5157 &   &   & 68 &   & 46 & 35 &   &   & \\
1930 & Lucifer & 32 & 17 & 211 & $-$19 & 13.0536 & 6 & 1 & 106 &   & 43 & 66 &   &   & \\
2156 & Kate & 49 & 74 &   &   & 5.62215 & 4 & 1 &   &   & 44 &   &   &   & \\
3678 & Mongmanwai & 125 & $-$65 &   &   & 4.18297 & 2 & 1 &   &   & 103 &   & 31 &   & \\
4483 & Petofi & 107 & 40 &   &   & 4.33299 & 3 & 1 &   &   & 36 &   &   &   & \\
\hline
\end{longtable}
\tablefoot{
For each asteroid, the table gives also the number of dense lightcurves $N_{\mathrm{lc}}$ observed during $N_{\mathrm{app}}$ apparitions and the number of sparse data points for the corresponding observatory: $N_{\mathrm{689}}$, $N_{\mathrm{699}}$, $N_{\mathrm{703}}$, $N_{\mathrm{E12}}$, $N_{\mathrm{G96}}$, $N_{\mathrm{950}}$ and $N_{\mathrm{Hip}}$. Pole solutions preferred by asteroid occultation measurements \citep{Durech2011s} are emphasized by a bold font.
}
}

\begin{table*}
\caption{\label{partials_tab}List of partial models derived from combined data sets.}
\begin{tabular}{r@{\,\,\,}l rrD{.}{.}{6}rrrrrrrrrrc}
\hline 
 \multicolumn{2}{c} {Asteroid} & \multicolumn{1}{c} {$\beta$} & \multicolumn{1}{c} {$\Delta$} & \multicolumn{1}{c} {P} & N$_{\mathrm{lc}}$ & N$_{\mathrm{app}}$  & N$_{\mathrm{689}}$ & N$_{\mathrm{699}}$ & N$_{\mathrm{703}}$ & N$_{\mathrm{E12}}$ & N$_{\mathrm{G96}}$ & N$_{\mathrm{950}}$ & N$_{\mathrm{Hip}}$\\
\multicolumn{2}{l} { } & [deg] & [deg] & \multicolumn{1}{c} {[hours]} &  &  &  &  &  &  &  &  &  & \\
\hline\hline 
163 & Erigone & $-$60 & 14 & 16.1403 & 3 & 1 & 168 &   & 72 &   &   &   & \\
187 & Lamberta & $-$58 & 9 & 10.66703 & 9 & 2 & 159 &   & 52 & 53 &   &   & \\
233 & Asterope & 49 & 8 & 19.6981 & 13 & 3 & 184 &   & 80 &   &   & 165 & \\
272 & Antonia & $-$70 & 6 & 3.85480 & 5 & 1 & 109 &   & 60 &   & 36 &   & \\
281 & Lucretia & $-$54 & 11 & 4.349710 & 6 & 3 & 123 & 30 & 62 &   &   &   & \\
313 & Chaldaea & 33 & 18 & 8.38992 & 9 & 3 & 176 &   & 80 &   &   &   & \\
390 & Alma & $-$60 & 22 & 3.74116 & 2 & 1 & 109 &   &   & 34 &   &   & \\
510 & Mabella & $-$59 & 12 & 19.4304 & 6 & 2 & 145 &   & 60 &   &   &   & \\
550 & Senta & $-$63 & 13 & 20.5726 & 9 & 1 & 151 &   &   & 61 &   &   & \\
622 & Esther & $-$61 & 9 & 47.5042 & 5 & 1 & 120 &   & 60 &   &   &   & \\
692 & Hippodamia & $-$52 & 25 & 8.99690 & 3 & 1 & 114 &   & 78 & 32 &   &   & \\
733 & Mocia & 36 & 16 & 11.37611 & 2 & 1 & 175 &   & 44 &   &   &   & \\
746 & Marlu & $-$54 & 18 & 7.78887 & 3 & 1 & 133 &   & 47 & 34 &   &   & \\
784 & Pickeringia & 58 & 15 & 13.1699 & 1 & 1 & 188 &   & 67 & 32 &   &   & \\
823 & Sisigambis & 57 & 9 & 146.58 & 8 & 1 & 123 &   & 90 &   &   &   & \\
877 & Walkure & 53 & 12 & 17.4217 & 3 & 1 & 141 & 45 & 104 &   & 32 &   & \\
899 & Jokaste & $-$58 & 19 & 6.24812 & 3 & 1 & 140 &   &   & 43 &   &   & \\
1010 & Marlene & 46 & 7 & 31.066 & 8 & 1 & 104 &   & 52 &   &   &   & \\
1103 & Sequoia & $-$48 & 19 & 3.037977 & 2 & 1 & 111 &   & 36 & 30 &   &   & \\
1185 & Nikko & 46 & 12 & 3.78615 & 3 & 1 & 91 &   & 46 & 32 &   &   & \\
1188 & Gothlandia & $-$63 & 19 & 3.491820 & 2 & 1 & 129 & 33 & 67 & 41 &   &   & \\
1214 & Richilde & $-$59 & 15 & 9.86687 & 4 & 1 & 101 &   & 78 &   &   &   & \\
1282 & Utopia & $-$39 & 21 & 13.6228 & 4 & 1 & 116 &   & 72 &   &   &   & \\
1350 & Rosselia & $-$58 & 13 & 8.14011 & 1 & 1 & 114 &   & 48 &   &   &   & \\
1368 & Numidia & $-$50 & 14 & 3.640740 & 3 & 1 & 129 &   & 47 &   &   &   & \\
1379 & Lomonosowa & $-$62 & 17 & 24.4845 & 2 & 1 & 96 &   & 100 &   &   &   & \\
1389 & Onnie & $-$56 & 10 & 23.0447 & 2 & 1 & 85 & 33 & 47 & 32 & 40 &   & \\
1665 & Gaby & 49 & 17 & 67.905 & 1 & 1 & 81 &   & 80 &   &   &   & \\
1719 & Jens & $-$56 & 19 & 5.87016 & 2 & 1 & 78 &   & 48 & 40 &   &   & \\
2001 & Einstein & $-$51 & 22 & 5.4850 & 2 & 1 &   &   & 84 &   &   &   & \\
\hline
\end{tabular}
\tablefoot{
For each asteroid, there is the mean ecliptic latitude $\beta$ of the pole direction and its dispersion $\Delta$, the other parameters have the same meaning as in Table \ref{models_tab}.
}
\end{table*}

\begin{table*}
\caption{\label{preliminary}List of improved asteroid models that were originally designated in DAMIT as ``preliminary``.}
\begin{tabular}{r@{\,\,\,}l rrrr D{.}{.}{6} rrrrrrrrrrc}
\hline 
\multicolumn{2}{c} {Asteroid} & \multicolumn{1}{c} {$\lambda_1$} & \multicolumn{1}{c} {$\beta_1$} & \multicolumn{1}{c} {$\lambda_2$} & \multicolumn{1}{c} {$\beta_2$} & \multicolumn{1}{c} {$P$} & $N_{\mathrm{lc}}$ & $N_{\mathrm{app}}$  & $N_{\mathrm{689}}$ & $N_{\mathrm{699}}$ & $N_{\mathrm{703}}$ & $N_{\mathrm{E12}}$ & $N_{\mathrm{G96}}$ & $N_{\mathrm{950}}$ & $N_{\mathrm{Hip}}$\\
\multicolumn{2}{l} { } & [deg] & [deg] & [deg] & [deg] & \multicolumn{1}{c} {[hours]} &  &  &  &  &  &  &  &  &\\ \hline\hline
73 & Klytia & 266 & 68 & 44 & 83 & 8.28307 & 21 & 7 & 131 & 36 & 98 & 47 &   &   & \\
82 & Alkmene & 164 & $-$28 & 349 & $-$33 & 13.00079 & 11 & 1 & 158 &   & 72 & 36 & 38 & 192 & \\
132 & Aethra & 326 & 67 &   &   & 5.16827 & 4 & 2 & 204 &   & 55 &   &   &   & \\
152 & Atala & \textbf{347} & \textbf{46} & 199 & 61 & 6.24472 & 2 & 1 & 101 &   & 32 &   &   &   & \\
277 & Elvira & 121 & $-$84 &   &   & 29.6922 & 22 & 5 & 142 &   & 36 & 51 &   &   & \\
278 & Paulina & 307 & 31 & 118 & 38 & 6.49387 & 3 & 1 & 195 &   & 51 &   &   &   & \\
311 & Claudia & 214 & 43 & 30 & 40 & 7.53138 & 23 & 6 & 114 & 33 & 108 &   & 40 &   & \\
484 & Pittsburghia & 70 & 46 &   &   & 10.64977 & 2 & 1 & 100 &   & 52 &   &   &   & \\
516 & Amherstia & 254 & 22 & 81 & 54 & 7.48431 & 5 & 3 & 162 &   & 32 &   &   &   & \\
534 & Nassovia & 66 & 41 & 252 & 42 & 9.46889 & 16 & 6 & 151 &   & 64 & 32 &   &   & \\
614 & Pia & 348 & 48 & 162 & 27 & 4.57872 & 2 & 1 & 121 &   & 78 &   &   &   & \\
714 & Ulula & 224 & $-$10 & 41 & $-$5 & 6.99837 & 9 & 2 & 177 &   & 67 &   &   &   & \\
770 & Bali & 70 & 50 & 262 & 45 & 5.81894 & 2 & 1 & 131 &   & 52 &   &   &   & \\
915 & Cosette & 350 & 56 & 189 & 61 & 4.469742 & 1 & 1 & 106 &   & 32 & 35 &   &   & \\
1012 & Sarema & 45 & 67 & 253 & 63 & 10.30708 & 2 & 1 & 74 &   & 42 &   &   &   & \\
1022 & Olympiada & 46 & 10 & 242 & 52 & 3.83359 & 5 & 2 & 107 &   & 91 &   &   &   & \\
1088 & Mitaka & 280 & $-$71 &   &   & 3.035378 & 1 & 1 & 104 &   & 39 & 41 &   &   & \\
1223 & Neckar & 252 & 28 & 69 & 30 & 7.82401 & 16 & 7 & 132 & 33 & 60 &   & 36 &   & \\
\hline
\end{tabular}
\tablefoot{
Pole solutions preferred by asteroid occultation measurements \citep{Durech2011s} are emphasized by a bold font.
}
\end{table*}

\begin{table*}
\caption{\label{references}Observations used for successful model determination that are not included in the UAPC.}
\begin{tabular}{r@{\,\,\,}l lc| r@{\,\,\,}l lc} \hline
 \multicolumn{2}{c} {Asteroid} & Date & Observer &  \multicolumn{2}{c}{Asteroid} & Date & Observer\\ \hline\hline
13 & Egeria & 2007.9$-$2009.3 & \citet{Pilcher2009a} &	808 & Merxia & 2003  1 26.8 & Casulli\tablefootmark{a}\\
14 & Irene & 2007.11$-$2009.5 & \citet{Pilcher2009a} &	 && 2003  1 28.8 & Casulli\tablefootmark{a} \\
 && 2008 2 3.7 & Polishook&	 && 2003  2 7.0 & Bernasconi\tablefootmark{a} \\
40 & Harmonia & 2008.12-2009.1 & \citet{Pilcher2009a} &	 && 2003  2 8.0 & Bernasconi\tablefootmark{a} \\
 && 2010.5$-$2010.6 & \citet{Pilcher2010a} &	832 & Karin & 2003.8$-$2003.9 & \citet{Yoshida2004} \\
68 & Leto & 2008 1 26.0 & Pilcher\tablefootmark{b} &	 && 2004.9$-$2004.9 & \citet{Ito2007} \\
 && 2008 1 30.0 & Pilcher\tablefootmark{b} &	899 & Jokaste & 2003.11$-$2003.12 & \citet{Stephens2004b} \\
 && 2008 2 6.1 & Pilcher\tablefootmark{b} &	1010 & Marlene & 2005.1$-$2005.3 & \citet{Warner2005} \\
264 & Libussa & 2005.2$-$2005.3 & \citet{Pilcher2006a} &	1022 & Olympiada & 1999.6$-$1999.6 & \citet{Warner2005a} \\
 && 2008.10$-$2008.12 & \citet{Pilcher2009b} &	1087 & Arabis & 2003  2 23.0 & Lehk\'y \\
272 & Antonia & 2007.12$-$2008.1 & \citet{Pilcher2008a} &	1140 & Crimea & 2005.4$-$2005.4 & \citet{Stephens2005a} \\
310 & Margarita & 2010.3$-$2010.5 & \citet{Pilcher2010b} &	1185 & Nikko & 2004.11$-$2004.11 & \citet{Stephens2005b} \\
390 & Alma & 2004.8$-$2004.8 & \citet{Stephens2005c} &	1282 & Utopia & 2000.11$-$2000.11 & \citet{Warner2011b} \\
400 & Ducrosa & 2005.1$-$2005.1 & \citet{Warner2005} &	1333 & Cevenola & 2002.2$-$2002.2 & \citet{Warner2002} \\
436 & Patricia & 2002.12$-$2003.1 & \citet{Warner2003} &	1635 & Bohrmann & 2003.9$-$2003.10 & \citet{Stephens2004a} \\
544 & Jetta & 2004.8$-$2004.8 & \citet{Stephens2005c} &	1659 & Punkaharju & 2000.11$-$2000.11 & \citet{Warner2011a} \\
573 & Recha & 2001.1$-$2001.1 & \citet{Warner2011b} &	1719 & Jens & 2000.9$-$2000.9 & \citet{Warner2011a} \\
714 & Ulula & 2005  9 23.8 & Henych &	1930 & Lucifer & 2003.10$-$2003.10 & \citet{Warner2005} \\
 && 2005  9 25.8 & Henych &	2001 & Einstein & 2004.12$-$2004.12 & \citet{Warner2005} \\
 && 2005  9 30.8 & Henych &	3678 & Mongmanwai & 2003.3$-$2003.3 & \citet{Stephens2003a} \\ 
 && 2005 10  1.8 & Henych &	&&&\\
 && 2005 10 10.8 & Henych &	&&&\\ \hline
\end{tabular}
\tablefoot{
\tablefoottext{a}{On line at \texttt{http://obswww.unige.ch/$\sim$behrend/page\_cou.html}}
\tablefoottext{b}{On line at \texttt{http://aslc-nm.org/Pilcher.html}}
}
\end{table*}

\end{document}